\documentclass[12pt,number,sort&compress,preprint]{elsarticle}
\usepackage{amsmath,amssymb}
\usepackage{mathtools}
\usepackage[utf8]{inputenc}
\usepackage[english]{babel}
\usepackage{csquotes}
\usepackage{graphicx}
\usepackage[binary-units]{siunitx}
\usepackage{booktabs,multicol} 
\usepackage[textsize=footnotesize,textwidth=1.9\linewidth]{todonotes}
\usepackage{caption}
\usepackage{fix-cm}
\usepackage{textcomp}
\usepackage{xspace,relsize}
\usepackage[version=4]{mhchem}
\usepackage{subcaption} 
\usepackage[symbol]{footmisc}
\usepackage{epstopdf} 
\usepackage{perpage} 
\usepackage{multirow} 
\usepackage[T1]{fontenc}
\usepackage{setspace}
\usepackage{pdfpages}

\usepackage{bm}

\usepackage{threeparttable}

\usepackage{etoolbox}

\usepackage[breaklinks=true, linkcolor=blue, citecolor=blue, colorlinks=true]{hyperref}
\usepackage[english,capitalise]{cleveref}
\crefformat{appendix}{#2#1#3}

\makeatletter
\@ifpackagelater{subcaption}{2017/01/01}
  {%
    \def\fullsubcapboxwidth/{1.0}%
  }
  {%
    \def\fullsubcapboxwidth/{0.49}%
  }%
\makeatother

\graphicspath{{./figures/}}

\DeclarePairedDelimiter{\abs}{\lvert}{\rvert}
\DeclarePairedDelimiter{\norm}{\lVert}{\rVert}
\DeclarePairedDelimiter{\paren}{(}{)}
\DeclarePairedDelimiter{\conc}{\lbrack}{\rbrack}

\makeatletter
\DeclareRobustCommand{\regmark}{\raisebox{1.13ex}{%
  \fontsize{.4\dimexpr\f@size pt}\z@\selectfont\textregistered}%
}
\makeatother

\DeclareSIUnit\atm{atm}

\newcommand{\Rplus}{\protect\hspace{-.1em}\protect\raisebox{.35ex}{\smaller{\smaller\textbf{+}}}}
\newcommand{\Cpp}{\mbox{C\Rplus\Rplus}\xspace}

\MakePerPage{footnote}

\journal{Computer Physics Communications}

\begin{document}

\begin{frontmatter}

\title{Accelerating reactive-flow simulations using vectorized chemistry integration}

\author[1]{Nicholas J.~Curtis}
\author[2]{Kyle E.~Niemeyer\corref{corr}}
\ead{kyle.niemeyer@oregonstate.edu}
\author[1]{Chih-Jen Sung}

\address[1]{Department of Mechanical Engineering, University of Connecticut, Storrs, CT 06269, USA}
\cortext[corr]{Corresponding author}
\address[2]{School of Mechanical, Industrial, and Manufacturing Engineering, Oregon State University, Corvallis, OR 97331, USA}

\begin{abstract}
The high cost of chemistry integration is a significant computational bottleneck for realistic 
reactive-flow simulations using operator splitting. Here we present a methodology to accelerate 
the solution of the chemical kinetic ordinary differential equations using single-instruction, 
multiple-data vector processing on CPUs using the OpenCL framework.
First, we compared several vectorized integration algorithms using chemical kinetic source terms 
and analytical Jacobians from the \texttt{pyJac} software against a widely used integration code, \texttt{CVODEs}.
Next, we extended the \texttt{OpenFOAM} computational fluid dynamics library to incorporate the vectorized solvers, 
and we compared the accuracy of a fourth-order linearly implicit integrator--both in vectorized 
form and a corresponding method native to \texttt{OpenFOAM}---with the community standard chemical kinetics 
library \texttt{Cantera}. We then applied our methodology to a variety of chemical kinetic models, 
turbulent intensities, and simulation scales to examine a range of engineering and scientific 
scale problems, including (pseudo) steady-state as well as time-dependent Reynolds-averaged 
Navier--Stokes simulations of the Sandia flame D and the Volvo Flygmotor bluff-body stabilized, 
premixed flame. Subsequently, we compared the performance of the vectorized and native \texttt{OpenFOAM} 
integrators over the studied models and simulations and found that our vectorized approach 
performs up to \numrange{33}{35}$\times$ faster than the native \texttt{OpenFOAM} solver with high accuracy.
\end{abstract}


\end{frontmatter}

\section{Introduction}

Developers of next-generation combustion devices have increasingly turned to new operating
regimes---e.g., low-temperature combustion~\cite{IMTENAN2014329}---to achieve higher efficiencies
and reduced emissions. Computational reactive-flow modeling has been used to rapidly develop 
these new combustion concepts into functional prototypes~\cite{WESTBROOK2005125}, but accurately 
predicting combustion processes at these regimes depends on realistic chemical kinetics models.
This places a large computational burden on reactive-flow modeling, since solving the ordinary 
differential equations (ODEs) governing chemical kinetics can dominate the overall cost of 
simulations. For example, in a study of \textit{n}-dodecane spray injection, a high-resolution 
large-eddy simulation (LES) using a 54-species skeletal model and up to \num{22} million 
cells took around \num{48000} CPU hours (or about 20 days of wall-clock time) to 
complete a single realization of just \SI{2}{\milli\second} after start of 
injection~\cite{Moiz2016123}. Large model size and high numerical stiffness
further increase computational cost, and unfortunately chemical kinetic models relevant 
to transportation or power-generation applications exhibit these characteristics~\cite{Lu:2009gh}.

Many strategies have been developed to reduce the cost of using realistic, detailed chemical kinetic
models in reactive-flow simulations~\cite{Lu:2009gh,Turanyi:2014aa}.
Skeletal reduction methods aim to remove unimportant species and reactions from a chemical kinetic 
model while maintaining fidelity to the base detailed 
model~\cite{Lu:2006bb,Pepiot-Desjardins:2008,Hiremath:2010jw,Niemeyer:2010bt}.
In addition, species with similar thermochemical properties may also be lumped 
together~\cite{Lu:2007,Ahmed:2007fa,Pepiot:2008kq}, time-scale methods can be used to reduce 
numerical stiffness~\cite{Maas:1992aa,Lam:1994ws,Lu:2001ve,Gou:2010}, and 
tabulation\slash interpolation methods reuse previously computed results to accelerate 
simulations~\cite{Pope:1997wu,Ren:2014cd,tdac}. Often, several of these techniques are combined 
to achieve better performance~\cite{Lu:2008bi,Niemeyer:2014,Niemeyer:2015wq}, or are even 
applied dynamically throughout a simulation to achieve greater local computational 
savings~\cite{Liang:2009,Yang:2013ip,Curtis:2015}.

In addition to methods that reduce or approximate the base chemical kinetic model, performance 
can be increased by improving the ODE integration algorithms used to solve the chemical-kinetic 
equations. Some researchers have developed new integration algorithms specifically for efficiently 
solving chemical kinetics~\cite{Mott2000,Gou:2010,hansen18}, while others have examined the 
performance of existing algorithms for use in reactive-flow
simulations~\cite{Shi:2012aa,stone2014comparison,Niemeyer:2014aa,IMREN20161,CurtisGPU:2017,stone2018}.
Further, several codes have recently been developed for analytically evaluating the chemical-kinetic 
Jacobian matrix~\cite{Niemeyer:2016aa,CURTIS2018186,HANSEN2018257}, which may be used to improve 
the performance of implicit ODE integration schemes~\cite{Lu:2009gh}.

In recent years, high-performance computing hardware has been evolving from a homogeneous, 
CPU-based paradigm to one dominated by heterogeneous processing architectures. 
In particular, single-instruction multiple-thread (SIMT) processors have become widely available, 
such as graphics processing units (GPUs), and many studies have leveraged 
their high floating-operation throughput to accelerate chemical-kinetics integration.
Early work used GPUs to evaluate chemical-kinetic source terms or factorize the Jacobian matrix, 
but found significant speedups only for large chemical kinetic models~\cite{Spafford:2010aa,Shi:2011aa}.
Later, several studies implemented GPU-based explicit integration techniques to achieve order-of-magnitude 
or more speedups when integrating non-stiff or moderately-stiff chemical 
kinetics~\cite{Niemeyer:2011aa,Shi:2012aa,Niemeyer:2014aa}. Further work followed that developed GPU-based 
implicit integration schemes, but these exhibited decreased performance with increasing numerical stiffness 
due to thread divergence~\cite{Stone:2013aa,Sewerin20151375,curtis2017investigation}.

In contrast, vectorization based on the single-instruction multiple-data (SIMD) paradigm, 
like that found on modern CPUs, has not been as extensively investigated for use in 
chemical-kinetic integration, though SIMD vectorization has been used for ODE integration 
in related contexts~\cite{kroshko2013efficient,Linford:2009,Linford:2011}.
Che et al.~\cite{CHE2018101} used loop-unrolling\slash rearrangement, aligned arrays, and 
OpenMP~\cite{dagum1998openmp} compilation directives (i.e., \texttt{\#pragma}s) to achieve SIMD 
vectorization for accelerating an LES solver on the Intel Xeon Phi Many Integrated Core (MIC) 
co-processor. Stone et al.~\cite{stone2018} used the OpenCL~\cite{stone2010opencl} 
framework to vectorize a linearly implicit Rosenbrock integration method~\cite{wanner1991solving} on the 
CPU, GPU, and MIC architectures. In addition, we developed the open-source platform \texttt{pyJac} that can generate 
vectorized, OpenCL-based chemical kinetic source term and Jacobian evaluation codes~\cite{CURTIS2018186}.

In this work, we describe a method to implement SIMD-based vectorized ODE integration 
methods---coupled with vectorized, analytical chemical kinetic Jacobian and source-term 
evaluations---for use in realistic reactive-flow simulations. Our primary focus is finding
the acceleration and performance scaling achievable by the vectorized methods, so we examine 
these in applications with different turbulent and chemical time scales, mesh resolutions, 
and chemical-kinetic model sizes to assess their performance in a variety of engineering and 
scientific contexts. To do so, we extended the open-source computational fluid dynamics 
(CFD) code \texttt{OpenFOAM}~\cite{openfoam_paper} to use the vectorized methods. 

Our goal here is to demonstrate the significant reduction in computational cost achievable 
by fully vectorized chemical-kinetic integration on CPUs without compromising accuracy, 
rather than to perform high-fidelity 
studies with the aim of investigating the combustion physics involved. Although the open-source 
nature of \texttt{OpenFOAM} is advantageous for these purposes---allowing modification to enable the 
vectorized ODE integration methods---it lacks some commonly used models (e.g., multi-component 
and mixture-averaged species transport~\cite{kee1986fortran})
and boundary conditions (e.g., Navier--Stokes Characteristic Boundary Conditions~\cite{BAUM1995247,OKONGO2002330}).
It has to be pointed out that
the impact of the interaction of errors from computational grid size, subgrid models, 
numerical schemes, and chemical kinetic model selection on the accuracy of reactive-flow 
simulations is not well understood. For instance, Cocks et al.~\cite{COCKS20153394} demonstrated 
that four different numerical schemes (as provided by industry, research, and open-source CFD codes) 
are unable to produce consistent reactive-flow LES solutions for the commonly used Volvo bluff-body 
stabilized flame~\cite{sjunnesson1991lda,sjunnesson1991validation,sjunnesson1992cars},
even when using the same computational grid, subgrid, and chemistry models.
Similarly, Rochette et al.~\cite{ROCHETTE2018417} found that the numerical scheme and chemical kinetic 
model have a large effect on the agreement of the solution with experimental data used in a 
reactive-flow LES solution of the Volvo bluff-body stabilized flame problem.
The focus of this work is not model validation, so we will use only built-in \texttt{OpenFOAM} numerical schemes, 
subgrid models, and boundary conditions---substituting in our vectorized ODE solver---to simulate 
selected reactive-flow problems, demonstrating that the techniques developed here can further 
aid the development of standard models in \texttt{OpenFOAM}.

The rest of this article is organized as follows.
\cref{sbb:methods} describes the numerical methods and software used in this study.
Subsequently, \cref{sbb:valid} first verifies the vectorized solvers against a typical implicit 
integration algorithm, then confirms their coupling to the CFD code \texttt{OpenFOAM}~\cite{openfoam_paper}.
Then, \cref{S:results} presents two case studies that compare the performance of the vectorized 
solvers with those built into \texttt{OpenFOAM}: 
Sandia flame D~\cite{BARLOW19981087,BARLOW2005433,SCHNEIDER2003185} in \cref{S:sandia_flamed} 
and the Volvo Flygmotor bluff-body 
flame~\cite{sjunnesson1991lda,sjunnesson1991validation,sjunnesson1992cars} in \cref{S:volvo}.
Finally, in \cref{sbb:conc} we identify and discuss directions for future efforts.

\clearpage

\section{Numerical methods and software}
\label{sbb:methods}
\subsection{\texttt{pyJac} code-generation platform}
\label{s:pyjac_openfoam}

\texttt{pyJac}~\cite{CURTIS2018186} is an open-source software package that generates code for 
evaluating the chemical-kinetic source terms and analytical Jacobian matrix for a variety of 
execution, data-ordering, and matrix-format patterns. For full details, we direct 
readers to our prior work focusing on \texttt{pyJac}~\cite{Niemeyer:2016aa,CURTIS2018186}; 
however, here we highlight key points relevant to this work.
When using the constant-pressure assumption\footnote{In this context, ``constant pressure'' 
refers to the solution of chemical kinetics within a reaction sub-step of an 
operator-splitting scheme, rather than a general constant-pressure reactive-flow simulation.}, 
the resulting thermochemical state vector in \texttt{pyJac} is
\begin{equation}
\label{e:state_vec}
 \Phi = \left\{T, V, n_1, n_2, \ldots, n_{N_{\text{sp}} - 1}\right\} \;,
\end{equation}
where $T$ is the temperature of the gas mixture, $V$ is the volume, $n_i$ is the moles of species $i$, 
and $N_{\text{sp}}$ is the number of species in the model.
The last species in the model is typically the bath gas---\ce{N2} in this study---and 
omitted from the state vector, as it is calculated implicitly via the ideal gas equation of state:
\begin{equation}
 n = \frac{PV}{\mathcal{R}T} = \sum_{i=1}^{N_{\text{sp}}}{n_i} \;,
\end{equation}
where $n$ is the total number of moles of the gas mixture and $\mathcal{R}$ is the universal gas constant.
The moles of the last species in the model $n_{N_{\text{sp}}}$ is then calculated as
\begin{equation}
 \label{e:n_nsp}
 n_{N_{\text{sp}}} = \frac{PV}{\mathcal{R}T} - \sum_{i=1}^{N_{\text{sp}}-1}{n_i} \;.
\end{equation}
This formulation explicitly conserves mass in \texttt{pyJac}, and also ensures that the 
system of equations is not overconstrained~\cite{HANSEN2018257}.

Given a thermochemical state vector, \texttt{pyJac} can evaluate the chemical kinetic source rates
\begin{equation}
 \label{e:dphidt}
 \frac{\text{d} \Phi }{\text{d} t } = f\left(\Phi\right) = 
 \left\{\frac{\text{d} T}{\text{d} t},\frac{\text{d} V}{\text{d} t}, 
 \frac{\text{d} n_1}{\text{d} t}, \frac{\text{d} n_2}{\text{d} t}, 
 \ldots, \frac{\text{d} n_{N_{\text{sp}} - 1}}{\text{d} t} \right\} \;,
\end{equation}
which form the autonomous chemical kinetic ODEs.
In addition, \texttt{pyJac} can calculate the analytical chemical kinetic Jacobian
\begin{equation}
 \label{e:jac}
 \mathcal{J}_{i,j} = \frac{\partial f_i}{\partial \Phi_j},\qquad i,j=1, \ldots, N_{\text{sp}} + 1 \;.
\end{equation}

Although \texttt{pyJac} can evaluate the Jacobian in a sparse-matrix format (e.g., compressed row storage), 
we used a dense-matrix format to simplify implementing linear-algebra operations; 
extending this effort to use a sparse matrix is a goal for future studies.
We used \texttt{pyJac} to generate an explicit ``shallow''-vectorization, as described 
in~\cref{s:bbles_opencl}, using a vector width of eight double-precision floating-point 
operations\footnote{If the OpenCL vector width is longer than that implemented on the 
underlying hardware, the vector operation is implicitly converted to multiple smaller 
vector operations, similar to loop-unrolling optimizations.}, and ``C'' (or row-major) data ordering.

\subsection{OpenCL and vectorization}
\label{s:bbles_opencl}

The parallel programming standard OpenCL~\cite{stone2010opencl} provides a common interface 
to execute vectorized code on a variety of different platforms (e.g., CPU, GPU, MIC).
For a detailed overview of different vectorization patterns and their applications to different 
hardware platforms for integrating chemical kinetic ODEs, we refer the reader to past 
works~\cite{CURTIS2018186,curtis2017investigation,Stone:2013aa,Bauer:2014}.
Here we will discuss only the so-called ``shallow''-vectorization pattern for SIMD 
processors (e.g., CPUs) using OpenCL.

This method groups together the thermochemical state vectors of several chemical kinetic 
ODE systems:
\begin{align}
 \Phi_{\text{vec}} = \Bigl\{& \left\{\Phi_{1, 1}, \Phi_{2, 1}, \ldots, \Phi_{N_v, 1}\right\}, \nonumber \\
                &\left\{\Phi_{1, 2}, \Phi_{2, 2}, \ldots, \Phi_{N_v, 2}\right\}, \ldots, \nonumber \\
		            &\left\{\Phi_{1, N_{\text{sp}}+1}, \Phi_{2, N_{\text{sp}}+1}, \ldots, \Phi_{N_v, N_{\text{sp}}+1}\right\} \Bigr\} \;,
\end{align}
where $N_v$ is the number of elements in the vector, also known as the vector width, 
and $\Phi_{j, i}$ is the $i$th component of the thermochemical state vector (\cref{e:state_vec}) 
for the $j$th state.

These modified state vectors can be loaded into OpenCL vector data types, e.g., \texttt{double8}, 
allowing floating-point math operations to be performed concurrently over the vector 
(by specialized vector processors present on all modern CPUs), thus accelerating computations.
The OpenCL runtime (e.g., as supplied by Intel~\cite{intelopencl:2018}) then
transforms the OpenCL code into vectorized operations using the vector-instruction set present 
on the device.

\subsection{\texttt{accelerInt} ODE integration library}
\label{s:accel}

Efficiently and accurately integrating the autonomous chemical kinetic ODEs is critical to 
simulating reactive flows. Integration advances the source rates---described 
in~\cref{s:pyjac_openfoam}---from an initial time $t_i$ to a final time $t_f$:
\begin{equation}
 \frac{\text{d} \Phi }{\text{d} t } = f\left(\Phi\right), \quad t_i \le t \le t_f \;.
\end{equation}

We have implemented several previously developed vectorized OpenCL-based ODE integration 
methods~\cite{stone2018} in the \texttt{accelerInt} software library~\cite{curtis2017investigation}: 
a fourth-order explicit Runge--Kutta method and several third- and fourth-order linearly 
implicit Rosenbrock methods~\cite{SANDU19973459,FATODE,wanner1991solving}. 
\Cref{T:solvers} lists the OpenCL-based solvers available 
in \texttt{accelerInt}, as well as their orders, solver types, and references.

\begin{table}[htbp]
\captionsetup{width=0.8\textwidth}
\centering
 \begin{tabular}{@{}l c l l l@{}}
 \toprule
 Solver name  & Order & Solver type       & Reference(s) & Short name \\
 \midrule
 Rosenbrock   & 3     & Linearly implicit & \cite{SANDU19973459,FATODE} & \texttt{ROS3} \\
 Rosenbrock   & 4     & Linearly implicit &\cite{wanner1991solving} & \texttt{ROS4} \\
 RODAS        & 3     & Linearly implicit &\cite{SANDU19973459,FATODE} & \texttt{RODAS3} \\
 RODAS        & 4     & Linearly implicit &\cite{wanner1991solving} & \texttt{RODAS4} \\
 RKF45 	      & 4     & Explicit          &\cite{Hairer1993} & \texttt{RKF45} \\
 \bottomrule
 \end{tabular}
 \caption{List of vectorized OpenCL integration methods incorporated into \texttt{accelerInt}.}
 \label{T:solvers}
\end{table}

Runge--Kutta methods include both implicit and explicit solvers, and are widely used to solve 
systems of stiff and non-stiff ODEs. A Runge--Kutta method with $s$ stages may be written as
\begin{equation}
 \label{e:rk}
 \Phi\left(t_{n+1}\right) = \Phi\left(t_{n}\right) + \sum_{i=1}^{s} b_i \boldsymbol{k}_i \;,
\end{equation}
where each stage is computed as
\begin{equation}
 \label{e:rk_stage}
 \boldsymbol{k}_i = h \, f \left(\Phi\left(t_{n}\right) + \sum_{j=1}^{s} a_{ij}\boldsymbol{k}_j\right) \;,
\end{equation}
where $h$ is the time-step size, and $a_{ij}$ and $b_i$ are method 
coefficients that define the algorithm.

The explicit solver included in \texttt{accelerInt} is a five-stage, fourth-order-accurate embedded 
Runge--Kutta--Fehlberg method (RKF45)~\cite{Hairer1993}. Explicit Runge--Kutta methods are obtained 
when the coefficient matrix $a_{ij}$ in~\cref{e:rk_stage} is strictly lower triangular 
(i.e., $a_{ii} = 0$). While explicit integration methods are efficient for non-stiff problems, 
they are only conditionally stable and perform poorly for stiff problems where the step size 
$h$ is limited by stability concerns rather than the desired accuracy.

Implicit Runge--Kutta methods result from a fully populated coefficient matrix $a_{ij}$, 
resulting in a system of non-linear equations that are typically solved via Newton--Raphson iteration.
This technique requires the repeated evaluation and factorization of the chemical-kinetic 
Jacobian matrix $\mathcal{J}$ (see~\cref{e:jac}). As such, implicit methods cost more per 
integration step, but their improved stability typically makes them more efficient for 
solving stiff ODEs. Implicit methods commonly reuse the Jacobian matrix 
(and its factorization) for multiple time steps to reduce the computational overhead.

Rosenbrock (ROS and RODAS)\footnote{Here we adopt the naming convention of Hairer and Wanner~\cite{wanner1991solving}.} 
methods are more similar in structure to Runge--Kutta methods than to fully implicit techniques.
They solve a linearized form of~\cref{e:rk} and are therefore known as ``linearly implicit'' techniques.
An $s$-stage Rosenbrock method is formulated as
\begin{align}
 \label{e:ros}
 &\boldsymbol{k}_i = h f\left( \Phi\left(t_{n}\right) + \sum_{j=1}^{i-1} \alpha_{ij}\boldsymbol{k}_j \right) + h \mathcal{J} \sum_{j=1}^{i} \gamma_{ij}\boldsymbol{k}_j, \quad i = 1,\ldots,s \\
 &\Phi\left(t_{n + 1}\right) = \Phi\left(t_{n}\right) + \sum_{i=1}^{s} b_i \boldsymbol{k}_i \;,
\end{align}
where $\alpha_{ij}$, $\gamma_{ij}$, and $b_i$ are the method parameters.
Typically Rosenbrock methods are constructed with $\gamma_{11} = \ldots = \gamma_{ss} = \gamma$, a constant parameter, 
such that only one LU decomposition must be performed per time step~\cite{wanner1991solving}.

To avoid solving a linear system and performing matrix-vector multiplication at each stage of the 
method~\cite{wanner1991solving,Kaps1985},~\cref{e:ros} may be transformed:
\begin{align}
 &\left(\frac{1}{h \gamma_{ii}} \bm{I} - \mathcal{J}\right) \bm{u}_i = 
 f\left( \Phi\left(t_{n}\right) + \sum_{j=1}^{i-1} a_{ij}\bm{u}_j \right) + 
 \sum_{j=1}^{i} \frac{c_{ij}}{h} \bm{u}_j, \quad i = 1,\ldots,s \\
 &\Phi\left(t_{n + 1}\right) = \Phi\left(t_{n}\right) + \sum_{j=1}^{s} m_j \bm{u}_j \;,
\end{align}
where $\bm{\Gamma} = \left(\gamma_{ij}\right)$ is an intermediate matrix constructed from the $\gamma_{ij}$ values,
\begin{align}
 \boldsymbol{u}_i &= \sum_{j=1}^{i} \gamma_{ij} \bm{k}_j \;, \\
 a_{ij} &= \alpha_{ij} \bm{\Gamma}^{-1} \;, \\
 c_{ij} &= \gamma \bm{I} - \bm{\Gamma}^{-1} \;, \text{ and} \\
 m_{j} &= b_{j} \bm{\Gamma}^{-1} \;.
\end{align}

Directly using the Jacobian matrix in Rosenbrock solvers avoids the need for Newton iteration, making 
these methods particularly well-suited for SIMD and SIMT vectorization due to the low levels of 
divergence between vector lanes\slash threads~\cite{stone2014comparison}.
However, ROS solvers are formulated around using an exact (analytical) Jacobian, since a 
finite-difference Jacobian may impact the solver's order and 
convergence~\cite{wanner1991solving,Shampine:1982:IRM:355993.355994}.
Further, the Jacobian must now be evaluated\slash factorized at each step, adding to the cost 
per time step. W-methods may be a suitable technique to avoid these costs, since they are 
formulated around using an inexact Jacobian~\cite{wanner1991solving}; 
these should be investigated for vectorized ODE integration in the future.

These Runge--Kutta and Rosenbrock solvers, coupled with OpenCL source-rate\slash Jacobian evaluation code 
generated by the latest version of \texttt{pyJac}~\cite{CURTIS2018186}, form the basis of the accelerated ODE 
integration techniques used here.

\subsection{\texttt{OpenFOAM}}
\label{S:of_coupling}

\texttt{OpenFOAM}~\cite{openfoam_paper} is an open-source \Cpp library capable of solving a variety of continuum 
mechanics problems, and is designed to allow straightforward extension and implementation of 
custom solvers. In this work, we extend the \texttt{OpenFOAM} applications for solving reactive-flow CFD 
problems to incorporate the vectorized ODE integration techniques outlined in~\cref{s:accel}.
The reactive-flow solver \texttt{reactingFoam} supports using a variety of boundary conditions, 
models, solution techniques, and turbulence descriptions, including Reynolds-averaged Navier--Stokes 
(RANS) and LES simulations.
For comprehensive details on these models and their implementation in \texttt{OpenFOAM} we refer readers to 
dedicated articles on these topics~\cite{Lysenko2014b,Fureby2957,openfoam_paper,openfoam_manual}.
Here we focus on the models relevant to incorporating \texttt{accelerInt} into \texttt{OpenFOAM} for solving chemical kinetic ODEs.

\subsubsection{Chemistry solvers}

\texttt{OpenFOAM} has a number of built-in solvers for integrating the chemical kinetic ODEs, including 
\Cpp implementations of the third- and fourth-order linearly implicit Rosenbrock solvers 
in \texttt{accelerInt}. However, the implementations differ in a number of key aspects.
First, and most obviously, the difference in programming languages results in different execution 
patterns: serial evaluation in \texttt{OpenFOAM} vs.\ vectorized\slash batched integration in \texttt{accelerInt}.
Second, the \texttt{ROS4} solver in \texttt{OpenFOAM} uses a different set of method 
coefficients~\cite{Shampine:1982:IRM:355993.355994} than those used in \texttt{accelerInt}~\cite{wanner1991solving}.
Third, \texttt{OpenFOAM} \texttt{v6.x} includes an analytical Jacobian evaluation code\footnote{Previous versions 
of \texttt{OpenFOAM} used a semi-analytical approach, where most Jacobian values were computed analytically but 
the temperature derivatives were evaluated using finite differences. The fully analytical Jacobian 
was introduced on the \href{https://github.com/OpenFOAM/OpenFOAM-dev}{\texttt{OpenFOAM}\texttt{-Dev}} channel in June 2018, 
and is built into \texttt{OpenFOAM} \texttt{v6.x}.}, but it uses a different state vector composed of the 
species concentrations, temperature, and pressure (which is assumed constant, as in \texttt{pyJac}).
Finally, \texttt{OpenFOAM} does not employ the explicit mass-conservation formulation employed by \texttt{pyJac}, 
i.e., the concentration of the last species in the model is solved for directly in \texttt{OpenFOAM}.


We created a new chemistry model for \texttt{OpenFOAM} by extending the base class 
\texttt{BasicChemistryModel}: \texttt{BatchedChemistryModel}. This new model performs
the chemical-kinetic integration of the thermochemical state vectors for the domain 
(or sub-domain, in the case of runs parallelized with the message passing interface, MPI~\cite{MPI3}) 
in a single batched call to the \texttt{accelerInt} library, instead of evaluating them sequentially as in the base \texttt{OpenFOAM} code.
The \texttt{accelerInt} library returns the updated thermochemical state vectors for the domain, as well 
as the final internal integration time step taken for each cell in the domain.
The time-step values are used as an initial time step for the next ODE integration of this cell 
(as in the \texttt{BasicChemistryModel}\footnote{Technically, this re-use takes place in the 
\texttt{StandardChemistryModel} in \texttt{OpenFOAM}, which is the sub-class of the \texttt{BasicChemistryModel} 
corresponding to our own \texttt{BatchedChemistryModel} implementation.} used in the \texttt{OpenFOAM} code),
as well as to adaptively limit the overall CFD time-step size for certain \texttt{OpenFOAM} solvers (e.g., \texttt{chemFoam}).

\subsubsection{Turbulent combustion model}

\texttt{OpenFOAM} implements multiple turbulent combustion models for simulating turbulence-chemistry 
interactions in reactive-flow simulations, ranging from a simple infinitely fast chemistry model 
to the more-complex flame surface density formulation~\cite{WELLER1998899}.
We selected the eddy-dissipation concept (EDC)~\cite{magnussen2005eddy} as a 
robust and relatively computationally expensive combustion model to demonstrate the 
performance of the vectorized ODE solvers.
EDC is a commonly used technique for modeling turbulence-chemistry interaction, 
and has been applied to a large variety of combustion problems~\cite{en11071902,edc1,edc2}.
Conceptually, EDC is based upon the idea of the turbulent energy cascade~\cite{magnussen2005eddy} 
and involves both a fine-scale reactor and its non-reacting surroundings~\cite{en11071902}.
EDC models molecular mixing between the fine scales and their surroundings by mass transfer between the two.
When using a detailed chemical kinetic model, the fine-scale reactor is typically treated as a 
perfectly stirred reactor and solved to equilibrium, imposing significant computational 
overhead~\cite{magnussen1989modeling}.

The mean reaction rate for species $i$ in the EDC model~\cite{magnussen2005eddy,en11071902}, 
$\overline{R}_i$, is given by\footnote{Here we only consider Magnussen's 2005 EDC model~\cite{magnussen2005eddy}. 
\texttt{OpenFOAM} implements other versions of the EDC model (B{\"o}senhofer et al.~\cite{en11071902} 
provided a good overview of the available versions), but the version selected does not affect 
the derivation of $\overline{R}_i$, as the chemical kinetic model is only responsible 
for evaluating the thermodynamic components of~\cref{e:EDC_final}.}
\begin{equation}
\label{e:EDC}
\overline{R}_i = \frac{\overline{\rho}}{\tau^{*}} \frac{\gamma_{L}^2 \chi}{1 - \gamma_{L}^2 \chi}\left(\overline{Y}_i - Y_i^{*}\right) \;,
\end{equation}
where $\overline{\rho}$ is the mean fluid density; $\overline{Y}_i$ and $Y_i^{*}$ are the 
fluid mean and fine-structure mass fractions of species $i$, respectively; $\tau^{*}$ and 
$\gamma_L$ are the fine-structure residence time and dimensionless length fraction, respectively; 
and $\chi$ is the fraction of fine-structure regions that interact with the rest of the fluid, 
typically assumed to be unity~\cite{en11071902}.

By definition, for the concentration of species $i$ we have
\begin{align}
 \label{e:conc_form1}
 \conc{C_i} &= \rho \frac{Y_i}{W_i} = \frac{n_i}{V} \;, \nonumber \\
 \intertext{and}
 \rho Y_i &= W_i \frac{n_i}{V} \;,
\end{align}
where $W_i$ is the molecular weight of species $i$ and the volume $V$ is that of the 
corresponding CFD cell. Combining~\cref{e:conc_form1} with \cref{e:EDC} gives the 
mean reaction rate in terms of species moles (the species state variable used in \texttt{pyJac}):
\begin{equation}
 \label{e:EDC_final}
 \overline{R}_i = \frac{1}{\tau^{*}} \frac{\gamma_{L}^2 \chi}{1 - \gamma_{L}^2 \chi} \frac{W_i}{\overline{V}} \left(\overline{n}_i - n_i^{*}\right) \;.
\end{equation}
where $\overline{n}_i$ and $n_i^{*}$ are the mean and fine-structure moles of species 
$i$, respectively. Both $\overline{n}_{N_{\text{sp}}}$ and $n_{N_{\text{sp}}}^{*}$, the numbers of moles 
of the last species in the model in the mean and fine structures, respectively, 
are calculated using~\cref{e:n_nsp} to be consistent with \texttt{pyJac}.

\section{Verification}
\label{sbb:valid}

In this section, we will verify the accuracy of the vectorized ODE solvers in two contexts.
First, we will directly compare the vectorized solvers implemented in \texttt{accelerInt} against the 
commonly used high-order implicit solver \texttt{CVODEs}~\cite{sundials:2.7.0,Brown:1989vl,Hindmarsh:2005}, 
to verify their accuracy and examine their relative performance.
Then, we will compare the accuracy of the vectorized \texttt{ROS4} solver coupled to \texttt{OpenFOAM} (using the methods 
described in~\cref{S:of_coupling}) and the same \texttt{ROS4} solver natively implemented in \texttt{OpenFOAM} to a 
reference chemical kinetics code \texttt{Cantera}~\cite{cantera} for constant-pressure homogeneous 
ignition problems.

\subsection{\texttt{accelerInt} verification}
\label{s:accel_valid}

To verify the new solvers, we sampled \num{100000} thermochemical conditions from a previously 
generated database~\cite{Curtis2016:ch4}, created using constant-pressure partially stirred 
reactor simulations~\cite{Niemeyer:2016aa} with the GRI-Mech 3.0~\cite{smith_gri-mech_30} 
chemical kinetic model, which consists of \num{53} species and \num{325} reactions.
The database covers a pressure range of \SIrange{1}{25}{\atm} and a range of temperatures and 
compositions from a cold unburned \ce{CH4}\slash air mixture to states of ignition and equilibrium.
These conditions were integrated using \texttt{CVODEs}~\cite{sundials:2.7.0,Brown:1989vl,Hindmarsh:2005} 
with tight integration tolerances (absolute tolerance of $\num{e-20}$, relative tolerance of 
$\num{e-15}$) for a single global time step of $\SI{e-6}{\sec}$ to form a reference solution.

We then used the OpenCL solvers to integrate the same initial values to the same end time, 
varying the tolerances given to the adaptive time-stepping algorithm over \num{e-4}, \num{e-5}, 
\ldots, \num{e-15}. We set both the relative and absolute tolerances for the OpenCL solvers to this 
tolerance value for the verification effort; in general these can be (and often are) 
different, e.g., as in the computation of the reference \texttt{CVODEs} solution.

The error of each initial value problem (IVP) was then measured as
\begin{equation}
  \label{e:ivp_norm}
  \norm{E_j} = \norm*{\frac{\abs{y_{i,j}^{\circ}\paren{t} - y_{i, j}\paren{t}}}{\num{e-10} + 
  \num{e-6} \times \abs{y_{i,j}^{\circ}\paren{t}}}}_{2}\;,
\end{equation}
where $y_{i,j}^{\circ}\paren{t}$ is the $i$th component of the solution computed by \texttt{CVODEs} 
for the $j$th IVP and $y_{i, j}\paren{t}$ is the solution computed by the solver being tested.
The \textquote{tolerances} used for calculating the weighted reference solution components 
in~\cref{e:ivp_norm} are for normalization purposes only, and are selected solely because they 
are the tolerances that will be used for the reactive-flow test-case (\cref{S:bluffbody_reactive}).
The error over all IVPs was then calculated using the infinity norm
\begin{equation}
 \label{e:error_norm}
 \norm{E} = \norm{E_j}_{\infty} = \max_j |E_j| \;.
\end{equation}

\begin{figure}[htbp]
\centering
\captionsetup{width=0.9\textwidth}
\includegraphics[width=0.9\textwidth]{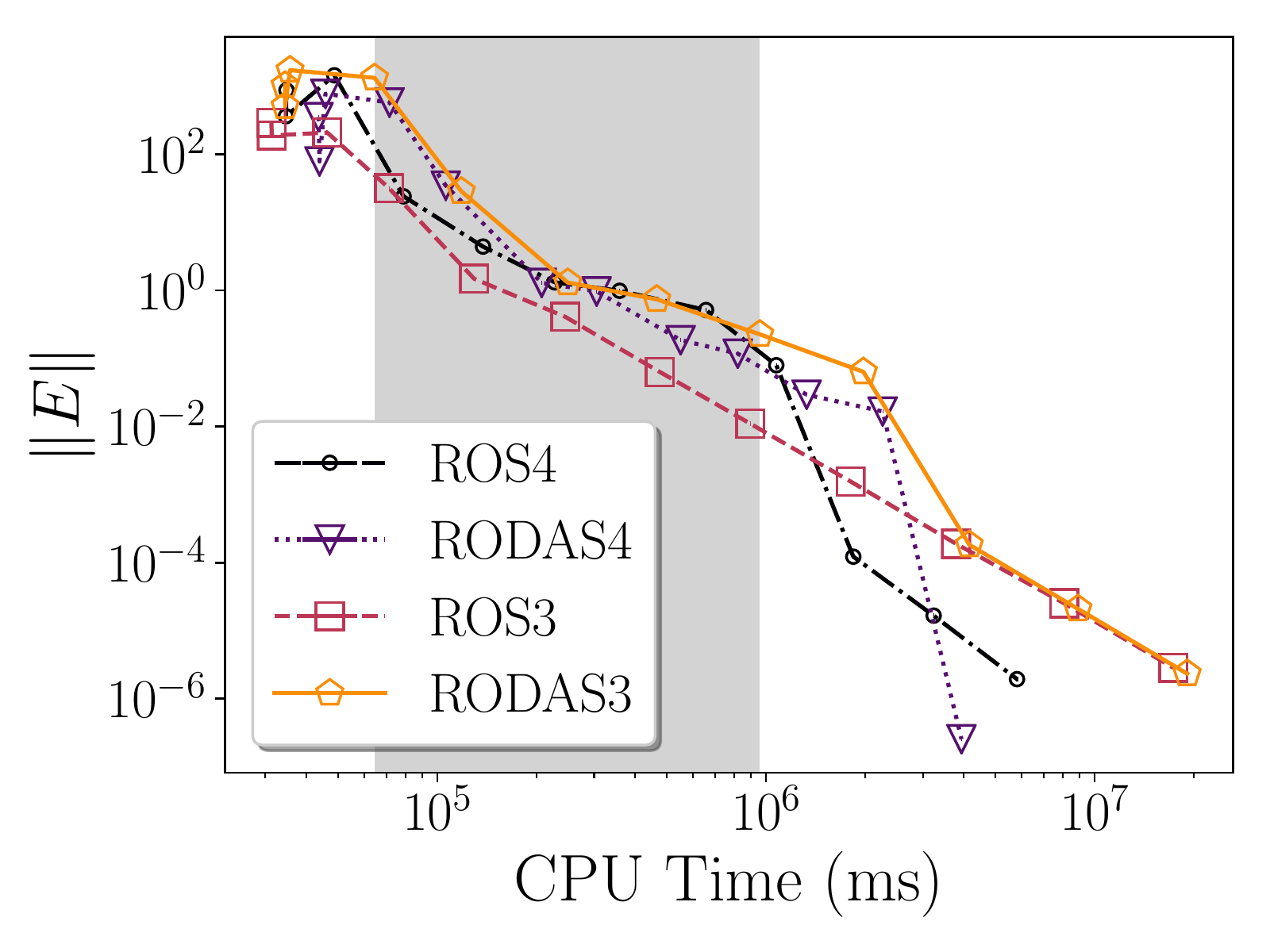}
\caption{Work-precision diagram for the implicit OpenCL solvers in \texttt{accelerInt}.
	 The vertical axis shows the error norm computed by~\cref{e:ivp_norm,e:error_norm}, 
   while the horizontal axis shows the mean runtime of the solver, measured in \si{\milli\second}.
	 The region marked in grey corresponds to ``intermediate'' tolerances, ranging from 
   \numrange{e-7}{e-11} (with equal absolute and relative tolerances in the adaptive 
   time-stepping algorithm).}
\label{F:accelerint_work_precision}
\end{figure}

The tested solvers used the same vectorization settings described in~\cref{s:pyjac_openfoam}.
\Cref{F:accelerint_work_precision} shows the work-precision diagram for the \texttt{accelerInt} solvers: 
the vertical axis shows the error (as measured by~\cref{e:ivp_norm,e:error_norm}) for the solvers 
over the various tolerances tested, while the horizontal axis shows the mean CPU runtime 
(averaged over five individual runs) on a single core of an Intel\regmark~Xeon\regmark~X5650 CPU 
(with SSE4.2 vector instructions), using \texttt{v16.1.1} of the Intel OpenCL runtime~\cite{intelopencl:2018}.
We omitted \texttt{RKF45} from this test, as the stiffness of the chemical kinetic ODEs 
caused prohibitive computational costs; this solver will be examined in future work for 
less-stiff problems.

For loose tolerances (\numrange{e-4}{e-6}), the tested solvers all exhibit similar performance 
and error. However, for intermediate tolerances (\numrange{e-7}{e-11}, marked in grey 
on~\cref{F:accelerint_work_precision}) the \texttt{ROS3} solver consistently has the lowest 
error compared with the reference solution; the fourth-order solvers (\texttt{ROS4}, \texttt{RODAS4}) 
tend to be slightly faster for a minor increase in error in this region.
The higher accuracy of the \texttt{ROS3} solver in this region is likely due to using a different 
set of method coefficients (see~\cref{T:solvers}). At strict tolerances, i.e., less than \num{e-11}, 
the \texttt{ROS4} solver is both faster and more accurate than the third-order methods, while the 
\texttt{RODAS4} solver is the most accurate for the strictest tolerance of \num{e-15}.

\subsection{Constant-pressure ignition in \texttt{OpenFOAM}}
\label{S:of_acc_conp}

\begin{figure}[htbp]
 \centering
 \subcaptionbox{%
  The predicted temperature traces at $P_0=\SI{25}{\atm}$, $T_0=\SI{850}{\kelvin}$, and \ce{CH4}\slash air of $\phi=1.5$. %
  The vertical dashed lines show the points at which the solution was sampled for error evaluation.\label{F:conp_temperature}%
  }[0.49\linewidth]{%
 \includegraphics[width=\fullsubcapboxwidth/\linewidth]{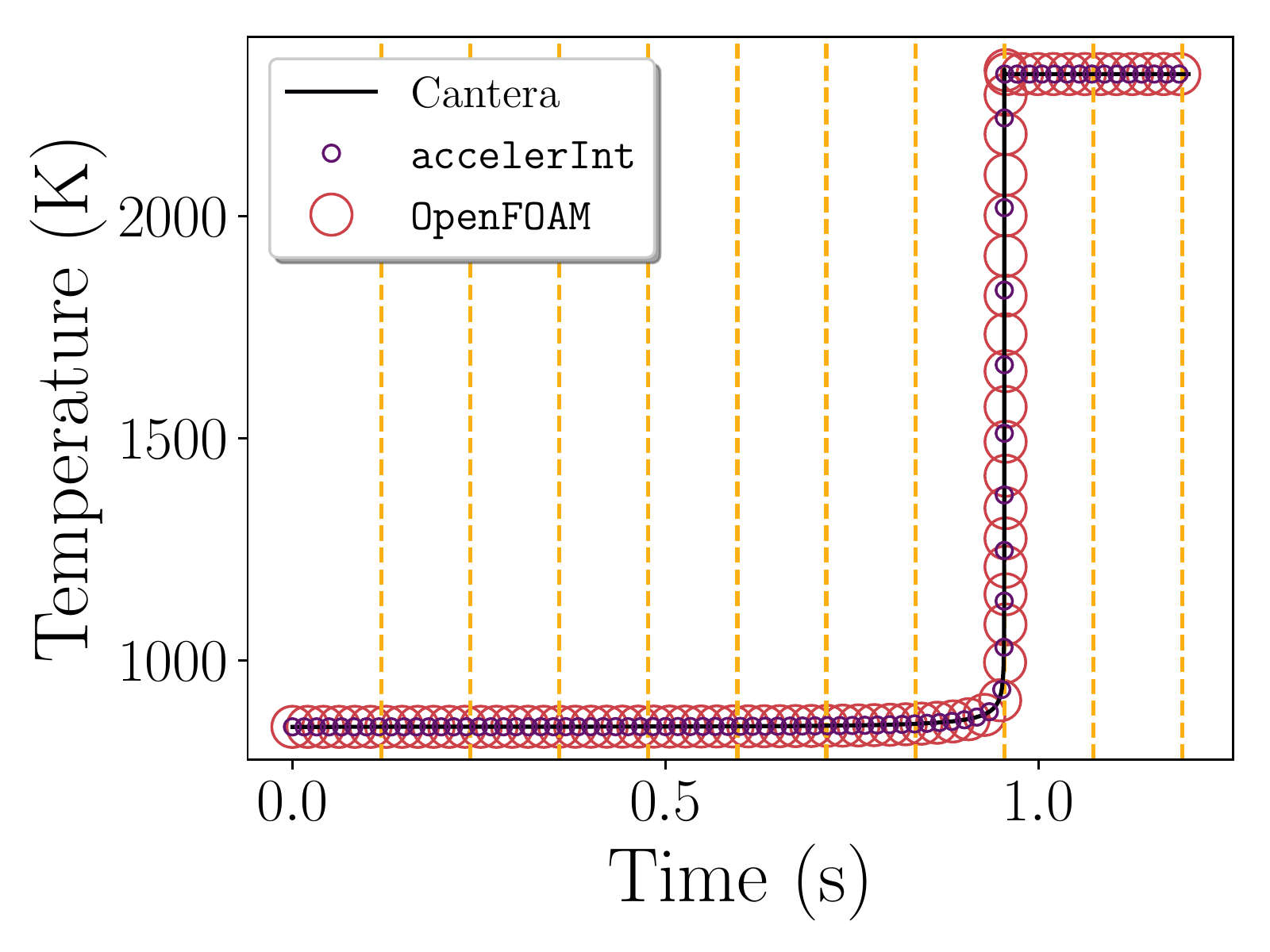}%
 }\hfill%
 \subcaptionbox{%
  The \ce{CH4} mass-fraction profiles predicted by the various solvers at $P_0=\SI{10}{\atm}$, $T_0=\SI{1100}{\kelvin}$, and \ce{CH4}\slash air of $\phi=1.0$%
  \label{F:conp_CH4_ign}%
  }[0.49\linewidth]{%
 \includegraphics[width=\fullsubcapboxwidth/\linewidth]{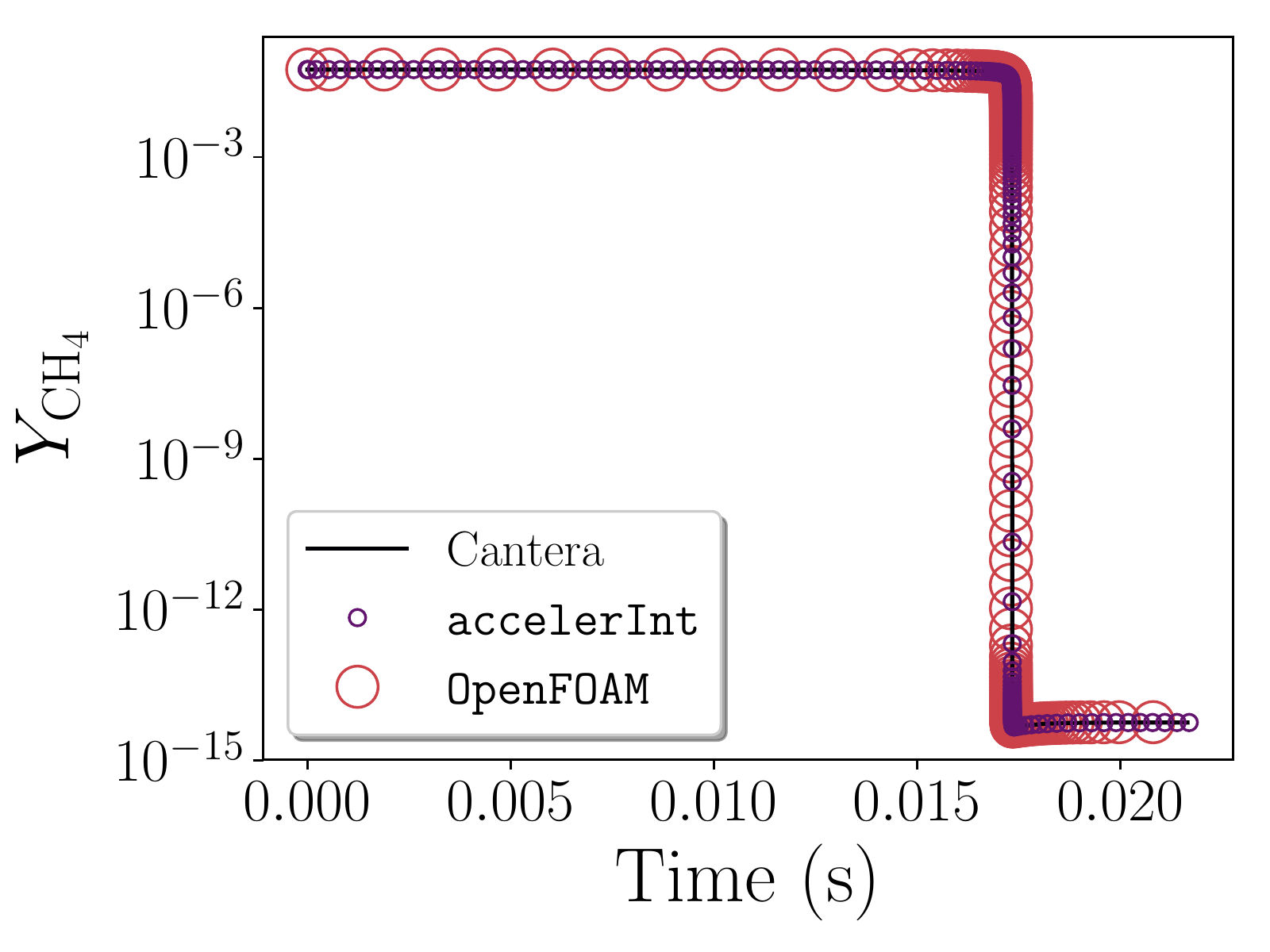}%
 }\\
 \subcaptionbox{%
  The \ce{OH} mass-fraction profiles predicted by the various solvers at $P_0=\SI{25}{\atm}$, $T_0=\SI{1500}{\kelvin}$, and \ce{CH4}\slash air of $\phi=0.5$%
  \label{F:conp_OH_ign}%
  }[0.49\linewidth]{%
 \includegraphics[width=\fullsubcapboxwidth/\linewidth]{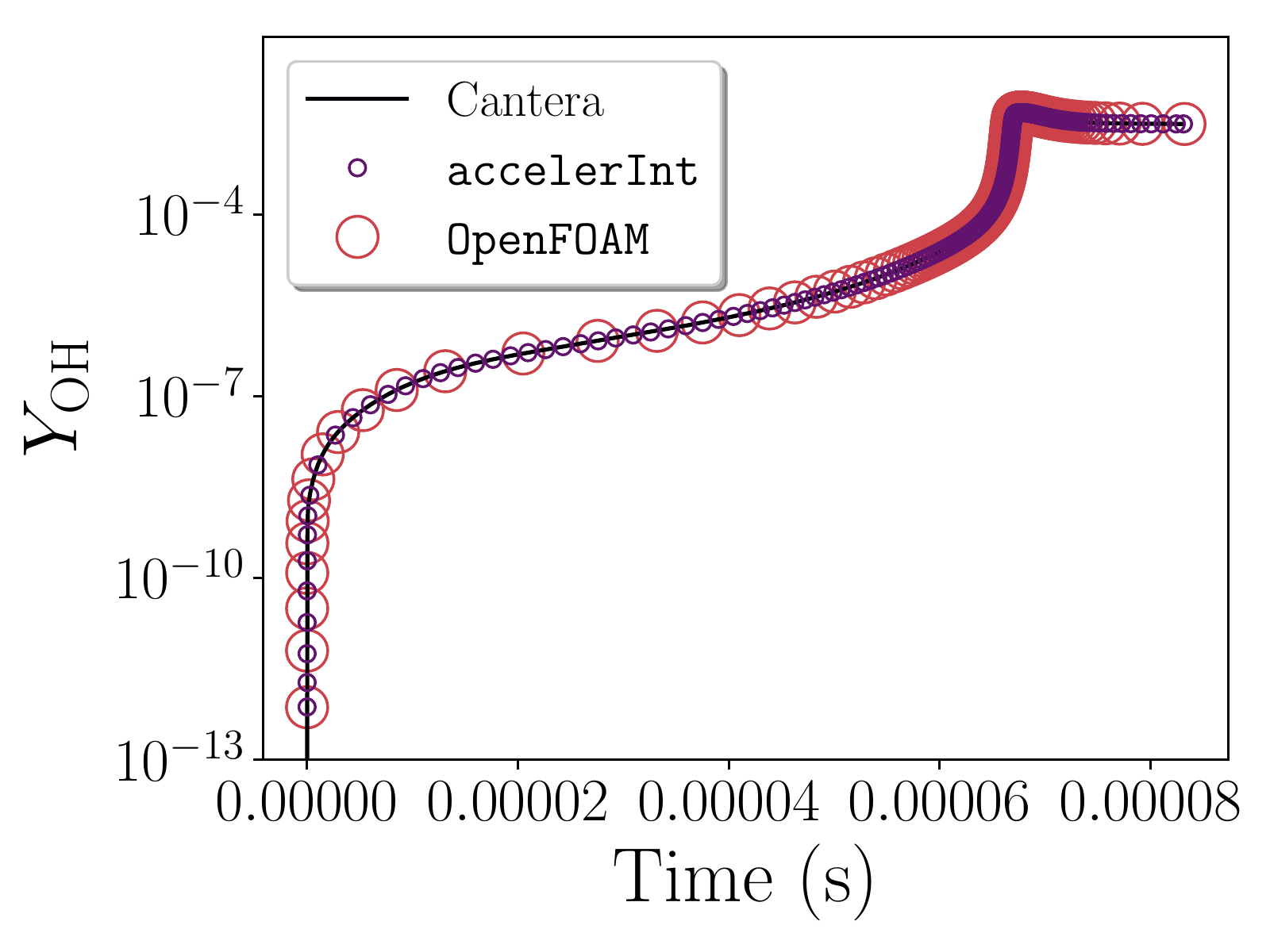}%
 }\hfill%
 \subcaptionbox{%
  A zoomed-in look at the \ce{NO} mass-fraction profiles predicted by the various solvers during the ignition event, for $P_0=\SI{25}{\atm}$, $T_0=\SI{850}{\kelvin}$, and \ce{CH4}\slash air of $\phi=1.0$.%
  \label{F:conp_NO_ign}%
  }[0.49\linewidth]{%
 \includegraphics[width=\fullsubcapboxwidth/\linewidth]{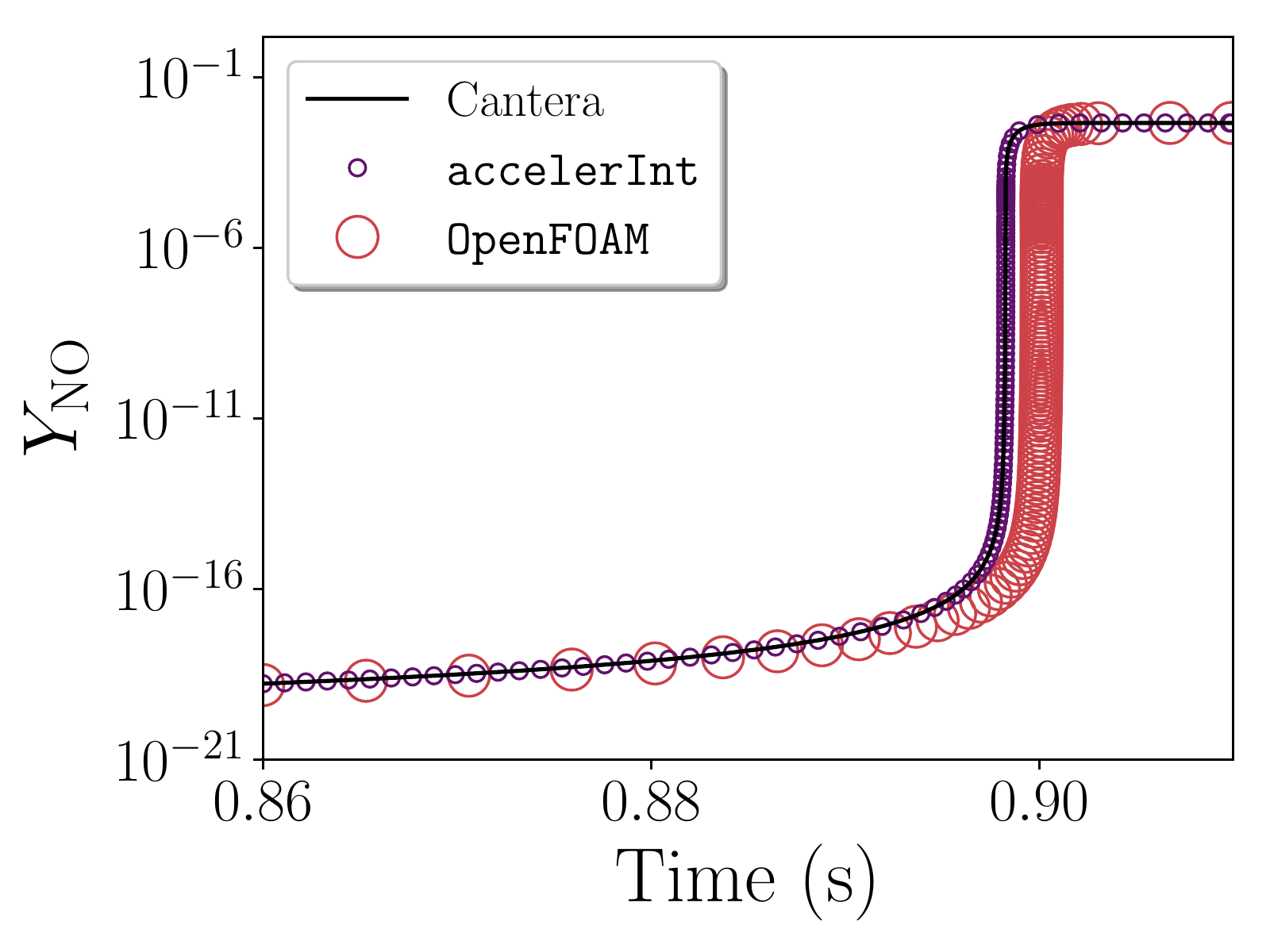}%
 }
 \caption{Comparison of constant-pressure homogeneous ignition problem with various solvers using GRI-Mech 3.0.
 The plotted points for the built-in \texttt{OpenFOAM} and coupled \texttt{accelerInt} \texttt{ROS4} solvers were thinned somewhat for visibility.}
 \label{F:conp_ign}
\end{figure}

To verify the coupling of the \texttt{accelerInt} solvers to the \texttt{OpenFOAM} chemistry model (see~\cref{S:of_coupling}), 
we ran a series of constant-pressure homogeneous ignition problems over 
initial temperatures of $T_0 =$ \SIlist{850;1100;1500}{\kelvin}, initial pressures of $P_0 =$ \SIlist{1;10;25}{\atm},
and equivalence ratios of $\phi =$ \numlist{0.5;1.0;1.5} in air, using the GRI-Mech 3.0 
chemical kinetic model~\cite{smith_gri-mech_30} and \ce{CH4} as the fuel.
These conditions cover low-, intermediate-, and high-temperature ignition, as well as lean, 
stoichiometric, and rich fuel-air mixtures to test the accuracy of the solvers over different 
chemical kinetic regimes.
To test the relative accuracy of the various solvers, the ignition problems were simulated to 
near chemical equilibrium (post-ignition) with both \texttt{accelerInt} and the built-in \texttt{OpenFOAM} integrators.
The values of the thermochemical-state vectors for each solver were sampled \num{10} times, 
equally distributed over the entire simulated time span (excluding the initial state); 
\cref{F:conp_temperature} shows an example of the time sampling.
We compared the sampled values of both approaches with a reference solution computed using
the open-source, community chemical kinetics code 
\texttt{Cantera}\footnote{\texttt{Cantera} internally uses \texttt{CVODEs} for ODE integration.}~\cite{cantera}.
In this example we used the fourth-order linearly implicit Rosenbrock solver (\texttt{ROS4}) in 
both \texttt{OpenFOAM} and \texttt{accelerInt}, and set the absolute and relative integration tolerances to 
\num{e-10} and \num{e-6}, respectively, for the \texttt{OpenFOAM} and \texttt{accelerInt} solvers, and to 
\num{e-20} and \num{e-15} for \texttt{Cantera}.


\Cref{F:conp_ign} compares the values of the temperature and species mass fractions of 
\ce{CH4}, \ce{OH}, and \ce{NO} for several initial conditions, showing that all three 
solvers agree qualitatively.
To quantify this comparison, the supremum and mean $L^2$ norms of the (filtered) relative 
error between the tested solvers and \texttt{Cantera} were calculated respectively as
\begin{align}
 \norm{E}_\infty &= \norm*{\frac{\abs{\Phi^{\circ}_{i}\left(t_j\right) - \Phi_{i}\left(t_j\right)}}{\abs{\num{1e-10} + \Phi_{i}^{\circ}\left(t_j\right)}}}_\infty  \label{e:max_rel_err} \\
 \norm{E}_\text{mean} &= \frac{1}{N_s \left(N_{\text{sp}} + 1\right)} \norm*{\frac{\abs{\Phi^{\circ}_{i}\left(t_j\right) - \Phi_{i}\left(t_j\right)}}{\abs{\num{1e-10} + \Phi_{i}^{\circ}\left(t_j\right)}}}_2 \label{e:mean_rel_err} \;,
\end{align}
where $N_s$ is the total number of sampled points, $N_{\text{sp}} + 1$ is the size of the 
thermochemical state vector used in \texttt{OpenFOAM} (i.e., the temperature and all species concentrations), 
$\Phi_{i}\left(t_j\right)$ is the $i$th entry of the state vector calculated by either the 
built-in \texttt{OpenFOAM} or coupled \texttt{accelerInt} \texttt{ROS4} solver at the $j$th sampled point, 
and $\Phi^{\circ}_{i}\left(t_j\right)$ the reference value calculated by \texttt{Cantera}.

\cref{t:conp_ign} shows the values of these norms for both solvers; the coupled \texttt{accelerInt} solver 
agrees much better with the reference solution computed by \texttt{Cantera} than the built-in \texttt{OpenFOAM} solver.
Although the computed error norms for \texttt{OpenFOAM} are several orders of magnitudes larger 
(i.e., $\mathcal{O}\left(10^5\right)$--$\mathcal{O}\left(10^6\right)$ versus 
$\mathcal{O}\left(10^{-1}\right)$--$\mathcal{O}\left(1\right)$ for \texttt{accelerInt}), 
this does not imply that the built-in \texttt{OpenFOAM} \texttt{ROS4} solver is grossly inaccurate; 
indeed, a visual comparison of the solutions over the entire simulation 
(\cref{F:conp_CH4_ign,F:conp_OH_ign,F:conp_temperature}) show no easily discernible discrepancies.
Instead, this difference largely results from more-accurate predictions of ignition delay time 
on the part of \texttt{accelerInt}. For instance,~\cref{F:conp_NO_ign} shows the predicted mass fraction of 
\ce{NO} during ignition for the stoichiometric case with an initial pressure of \SI{25}{\atm} and 
temperature of \SI{850}{\kelvin}. The \texttt{OpenFOAM} solver predicts ignition about \SI{1.8}{\milli\second}
later than either \texttt{accelerInt} or \texttt{Cantera}, a mere \SI{0.21}{$\percent$} difference.
Nonetheless, we conclude that the coupled \texttt{accelerInt} \texttt{ROS4} solver is more accurate than the 
corresponding built-in \texttt{OpenFOAM} implementation.

\begin{table}[htbp]
\captionsetup{width=0.75\textwidth}
\centering
\begin{tabular}{@{}l S[table-format=1.2e1, retain-zero-exponent=true] S[table-format=1.2e1, retain-zero-exponent=true]@{}}
\toprule
\multicolumn{1}{l}{Solver} & \multicolumn{1}{c}{$\norm{E}_\text{mean}$}  &  \multicolumn{1}{c}{$\norm{E}_\infty$} \\
\midrule
\texttt{OpenFOAM}      & \num{4.68e5}   & \num{9.49e6} \\
\texttt{accelerInt}   & \num{3.39e-1}  & \num{8.12e0} \\
\bottomrule
\end{tabular}
\caption{The filtered mean and supremum relative error norms comparing the computed solutions of 
the built-in \texttt{OpenFOAM} and coupled \texttt{accelerInt} \texttt{ROS4} solvers to those of \texttt{Cantera} for the homogeneous 
constant-pressure ignition problems.}
\label{t:conp_ign}
\end{table}

\section{Case studies}
\label{S:results}

After verifying the accuracy of the \texttt{accelerInt} solvers, we next 
compare the performance of the built-in \texttt{OpenFOAM} and 
\texttt{accelerInt} \texttt{ROS4} solvers on realistic reactive-flow simulations: 
the Sandia flame D and the Volvo Flygmotor bluff-body stabilized flame.
Our objective in both cases is to determine the potential speedup offered
by the vectorized solver.

\subsection{Sandia flame D}
\label{S:sandia_flamed}
\begin{figure}[htbp]
 \centering
 \begin{subfigure}[b]{0.6\textwidth}
 \centering
 \includegraphics[width=\linewidth]{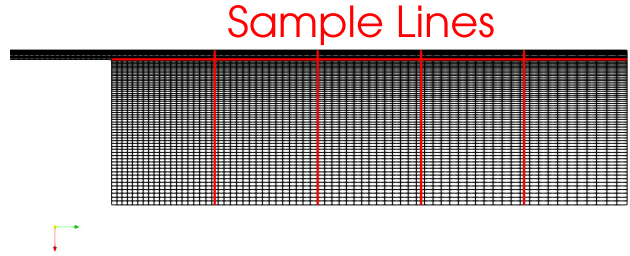}
 \caption{The mesh for the Sandia Flame D case. The red line denotes where we sampled the solution
  for validation.}
 \label{F:sandia_mesh}
 \end{subfigure}\\
 \begin{subfigure}[b]{0.6\textwidth}
 \centering
 \includegraphics[width=\linewidth]{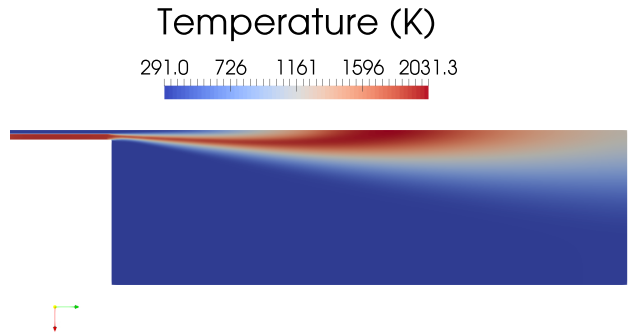}
 \caption{The steady-state temperature profile in~\si{\kelvin}, as solved by the base 
 \texttt{OpenFOAM} fourth-order Rosenbrock ODE integrator.}
 \label{F:sandia_steady_flame}
 \end{subfigure}\\
 \begin{subfigure}[b]{0.6\textwidth}
 \centering
 \includegraphics[width=\linewidth]{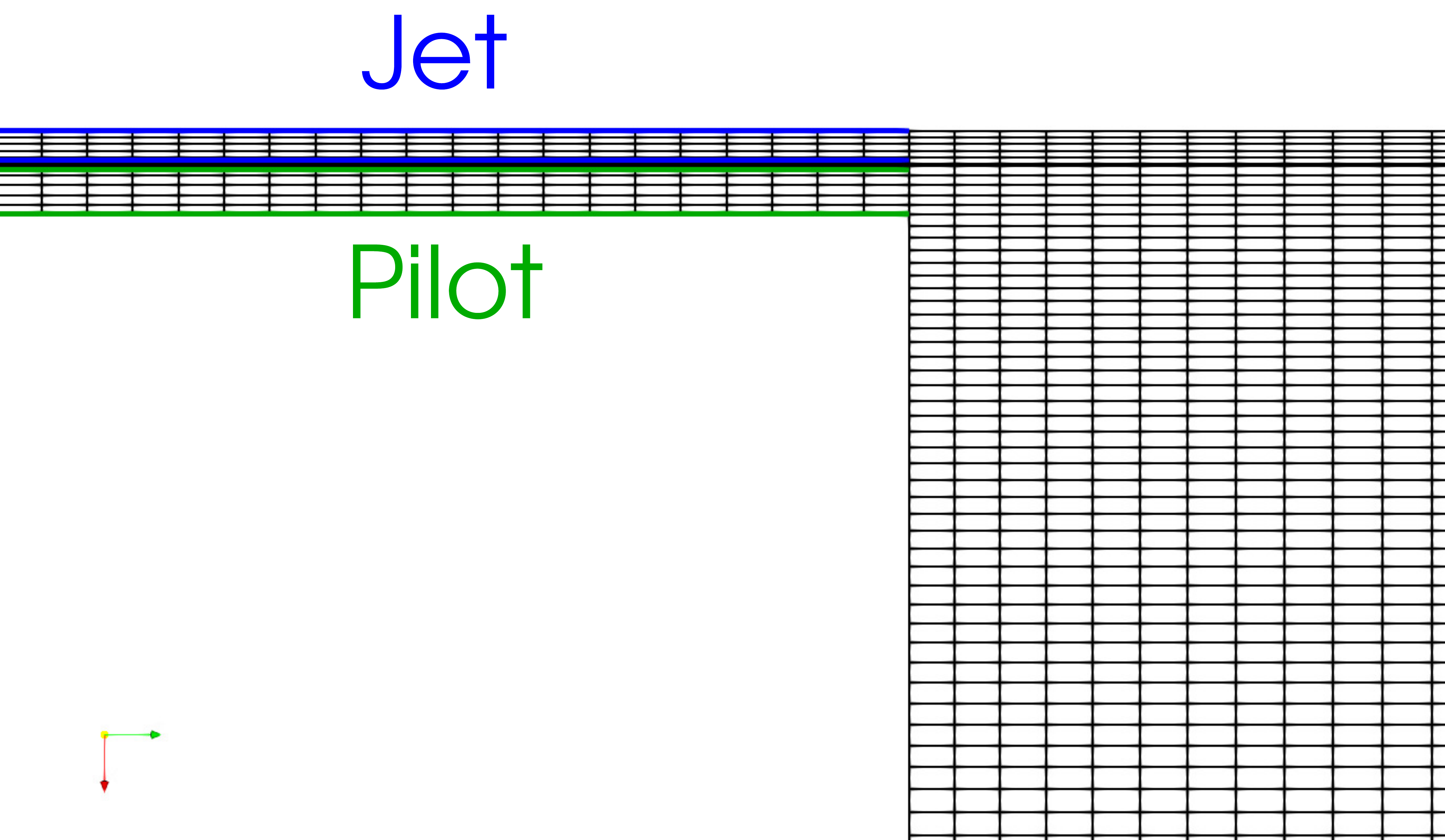}
 \caption{A close-up view of the inlet to the domain.
	  The jet and pilot flames are marked in blue and green, respectively.
	  The sample lines are marked in red, as in~\cref{F:sandia_mesh}}
 \label{F:sandia_schematic}
 \end{subfigure}
 \caption{The mesh and steady-state temperature profile of the Sandia Flame D case.}
\end{figure}

The Sandia Flame D~\cite{BARLOW19981087,BARLOW2005433,SCHNEIDER2003185} 
is a well-characterized piloted \ce{CH4}\slash air jet flame with a turbulent Reynolds number of 
$\text{Re}_{t} = 22000$. \texttt{OpenFOAM} \texttt{v5.x} (the 2017 release from the OpenFOAM 
foundation~\cite{openfoam_foundation}) and later include a case modeling this flame as a tutorial, 
providing a relatively inexpensive but more-realistic proving ground for the 
coupled \texttt{accelerInt} solvers. \cref{T:op_conds} and \cref{T:dimensions} list the operating conditions
and key dimensions of the case, and \cref{F:sandia_schematic} shows a schematic of the configuration 
superimposed over the simulation mesh.

\begin{table}[htbp]
  \centering
  \begin{threeparttable}
\begin{tabular}{@{}l l l l@{}}
  \toprule
  \multicolumn{1}{c}{} & \multicolumn{3}{c}{Source} \\
  \cmidrule(lr){2-4} & Jet           & Coflow       & Pilot \\
  \midrule
  Composition        & \ce{CH4}\slash\ce{Air}: 25\slash\SI{75}{\percent}\tnote{\textdagger} &  Dry air & Equil.\tnote{\textdaggerdbl} \\
  Velocity 	     & \SI{49.6}{\meter\per\second} & \SI{0.9}{\meter\per\second} & \SI{11.4}{\meter\per\second} \\
  Temperature 	     & \SI{294}{\kelvin}	    & \SI{291}{\kelvin}		  & \SI{1880}{\kelvin} \\
  Pressure	     & \SI{0.993}{\atm}		    & \SI{0.993}{\atm}		  & \SI{0.993}{\atm} \\
  \bottomrule
\end{tabular}
\caption{Operating conditions for the Sandia Flame D case.}
\begin{tablenotes}\footnotesize
\item[\textdagger] The jet composition is measured by percent volume.
\item[\textdaggerdbl] Equilibrium state of \ce{CH4}\slash\ce{air} at $\phi=0.7$.
\end{tablenotes} 
\label{T:op_conds}
\end{threeparttable}
\end{table}

\begin{table}[htbp]
  \centering
 \begin{tabular}{@{}l S[table-format=1.1,table-space-text-post=\si{\milli\meter}]@{}}
  \toprule
  Dimension          & \multicolumn{1}{c}{Value} \\
  \midrule
  $D_{\text{jet, inner}}$   & \SI{7.2}{\milli\meter} \\
  $D_{\text{pilot, inner}}$ & \SI{7.7}{\milli\meter} \\
  $D_{\text{pilot, outer}}$ & \SI{18.2}{\milli\meter} \\
  $D_{\text{wall, outer}}$  & \SI{18.9}{\milli\meter} \\
  Exit		     & \SI{30}{\milli\meter} $\times$ \SI{30}{\milli\meter} \\
  \bottomrule
\end{tabular}
\caption{Diameter of the jet, pilot, and wall, and exit-plane dimensions 
for the Sandia Flame D case.}
\label{T:dimensions}
\end{table}

The mesh for this case, pictured in~\cref{F:sandia_mesh}, is fully orthogonal with 
\num{5170} cells, ranging in size from roughly \SI{5}{\milli\meter} on a 
side to \SI{0.72}{\milli\meter} tall near the wall.
The solution domain is a thin three-dimensional wedge, with axi-symmetric boundary 
conditions on the front and back faces, and zero-gradient\slash total-pressure boundary 
conditions on the outlet. The case, as packaged with \texttt{OpenFOAM}, uses second-order interpolation 
and gradient schemes for all variables, but a strongly limited scheme (tending towards first-order) 
for divergence calculations. In addition, the simulation uses a standard $k$--$\varepsilon$ 
RANS model to model turbulence and a pseudo-transient, first-order time-stepping scheme 
to initially advance to a steady-state solution.

We modified the baseline case slightly to better suit the purposes of this study.
First, we substituted the full GRI-Mech 3.0 chemical kinetic model~\cite{smith_gri-mech_30} 
for the 36-species skeletal methane model originally used in \texttt{OpenFOAM}.
Second, the base \texttt{OpenFOAM} case uses a tabulated dynamic adaptive chemistry scheme~\cite{tdac,Ren:2014cd} 
to accelerate the solution process; we disabled this to directly compare the performance and 
accuracy of the ODE solvers. Finally, after reaching steady state, we switched the time-stepping scheme 
to the second-order implicit method, set the minimum reacting temperature to \SI{500}{\kelvin}, 
and ran the case for an additional \SI{10}{\milli\second} of simulated time using a CFD time step 
of $\Delta t = \SI{e-6}{\sec}$.

Using this case, we tested the performance of the various \texttt{OpenFOAM} and \texttt{accelerInt} solvers.
Our goal here is to demonstrate the performance enhancement achievable through vectorized 
chemical kinetic integration, using available Intel CPUs; though dated, 
the \texttt{AVX2} instruction set remains broadly relevant for CPUs that 
do not support the newer \texttt{AVX-512} set.
We ran the performance studies on \num{10} cores of an Intel E5-2690 V3 CPU, with 
\texttt{AVX2} vector instructions, \SI{128}{\giga\bit} of RAM, and \texttt{v16.1.1} of the 
Intel OpenCL runtime. We used \texttt{OpenFOAM} version \texttt{6.x} with \texttt{v2.1.0} of the 
OpenMPI library~\cite{open_mpi}, compiled with \texttt{gcc v5.4.0}~\cite{Stallman:2009:UGC:1593499}.
We instrumented the \texttt{reactingFoam} solver with the MPI profiling library 
\texttt{IPM v2.0.6}~\cite{skinner2005performance} by placing profiling sections 
(via~\texttt{MPI\_PControl}) around the calls to the turbulent combustion model, 
ODE integration, and other key parts of the CFD time step (e.g., convection evaluation).
Section S1 of the Supplemental Material compares in detail the predicted results
for the different solvers; here, we focus on the computational cost\slash performance.

\begin{figure}[htbp]
\centering
\includegraphics[width=0.75\textwidth]{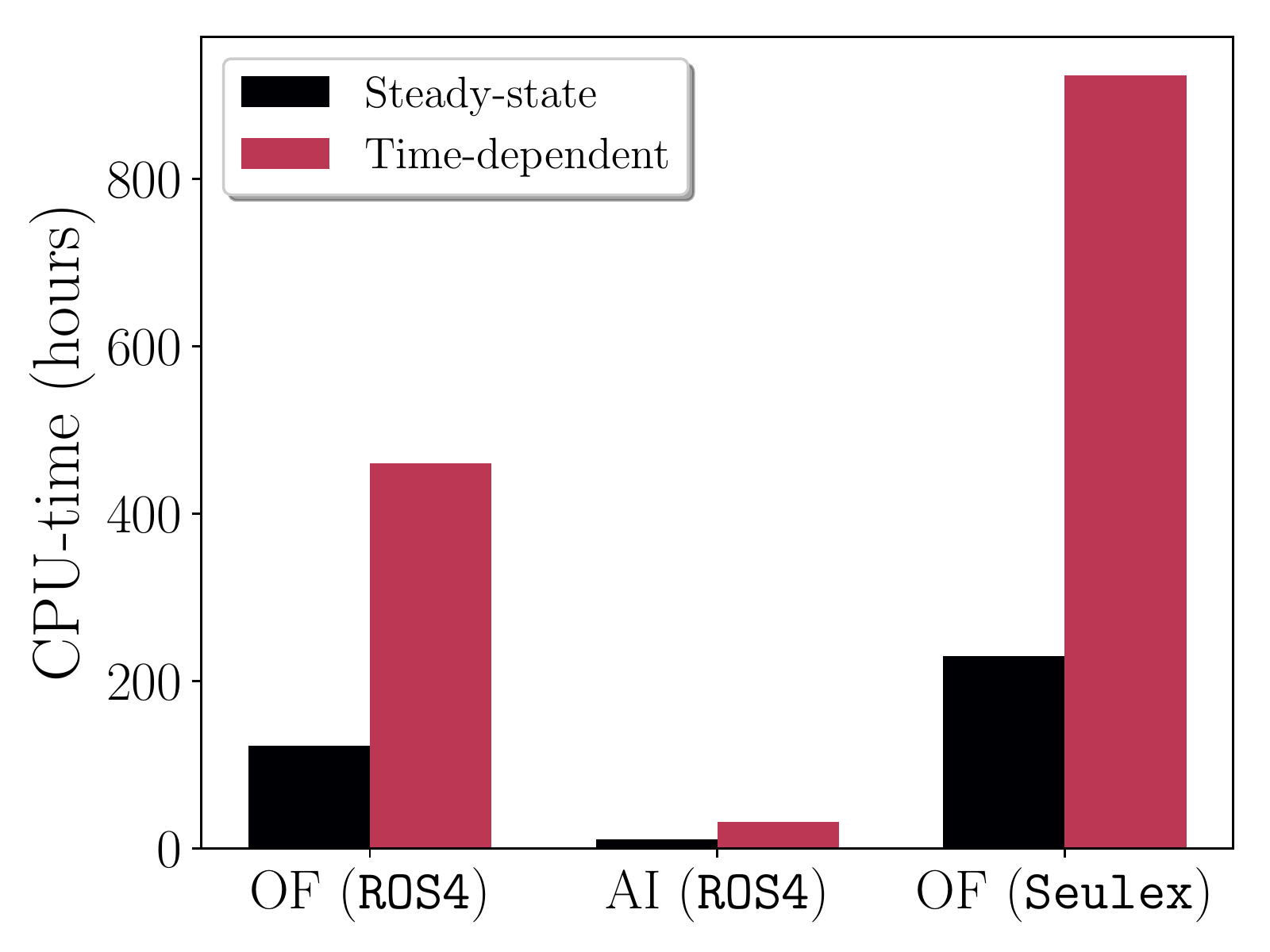}
\caption{Total CPU hours spent evaluating chemistry in the Sandia Flame D case.
AI is \texttt{accelerInt} and OF is \texttt{OpenFOAM}.}%
\label{F:sandia_walltime}
\end{figure}

\Cref{F:sandia_walltime} shows the total wall-clock execution time (over all \num{10} processors) 
spent by each solver integrating the chemical kinetic ODEs.
The \texttt{accelerInt} solver spends just 10.2 and \SI{31.1}{\hour} solving chemistry for the steady-state 
and time-dependent cases, respectively, while both the \texttt{OpenFOAM} solvers take over \SI{100}{\hour} in all cases.
The slowest integration method is the \texttt{OpenFOAM} \texttt{Seulex} solver, which takes over \SI{923}{\hour} 
to complete the time-dependent solution. The chemistry evaluation time includes 
time spent idle due to the poor load balancing of chemistry present in most \texttt{OpenFOAM} 
simulations\footnote{\texttt{OpenFOAM} uses a simple static decomposition of the domain, which is to say there 
is no chemistry load-balancing occurring.}.

\begin{table}[htbp]
  \centering
  \captionsetup{width=0.65\textwidth} 
  \begin{tabular}{@{} l S[table-format=1.2] S[table-format=1.2] S[table-format=1.2]}
  \toprule
		    & \multicolumn{3}{c}{Solver}\\
  \cmidrule(lr){2-4}& \multicolumn{1}{c}{AI (\texttt{ROS4})} & \multicolumn{1}{c}{OF (\texttt{ROS4})}  & \multicolumn{1}{c}{OF (\texttt{Seulex})}  \\
  \midrule
  Steady-state      & $\SI{12.0}{\times}$ 	    & \multicolumn{1}{c}{--}          & $\SI{0.53}{\times}$   \\
  Time-dependent    & $\SI{14.8}{\times}$ 	    & \multicolumn{1}{c}{--}          & $\SI{0.50}{\times}$    \\
  Chemistry time    & \SIrange{93.9}{95.8}{$\percent$} & \SIrange{99.3}{99.6}{$\percent$} & \SIrange{99.7}{99.9}{$\percent$} \\
  \bottomrule
  \end{tabular}
  \caption{Speedups of the chemistry solvers (AI is \texttt{accelerInt} and OF is \texttt{OpenFOAM}), normalized by the 
  \texttt{OpenFOAM} \texttt{ROS4} solver and the percent of total execution time spent integrating the 
  chemical-kinetic ODEs.}
  \label{T:sandia_perfomance}
\end{table}

\cref{T:sandia_perfomance} compares the speedups of the solvers, reported against the \texttt{OpenFOAM} \texttt{ROS4} 
solver as the baseline. The \texttt{accelerInt} \texttt{ROS4} solver performs \numrange{12.0}{14.8}$\times$ faster 
than the \texttt{OpenFOAM} equivalent. In addition, the \texttt{OpenFOAM} \texttt{Seulex} solver runs roughly 
2$\times$ slower than the \texttt{OpenFOAM} \texttt{ROS4} solver in both cases. Finally, the \texttt{accelerInt} solver 
spends \SIrange{94}{96}{$\percent$} in chemistry integration, while both \texttt{OpenFOAM} methods 
use over \SI{99}{$\percent$} of the runtime solving chemistry.

\subsection{Volvo bluff-body stabilized flame}
\label{S:volvo}

Next, we will use the Volvo Flygmotor bluff-body stabilized premixed 
flame experiment~\cite{sjunnesson1991lda,sjunnesson1991validation,sjunnesson1992cars} as a
second test case.
We simulated a reacting case, and used this to study the speedup of the 
coupled \texttt{accelerInt} solver compared to the built-in \texttt{OpenFOAM} method.

\subsubsection{Case description}
\label{sbb:case}
The Volvo Flygmotor bluff-body stabilized premixed flame experiments include many phenomena 
found in practical combustors, such as anchored flames, recirculation regions, and shear 
layers~\cite{sjunnesson1991lda,sjunnesson1991validation,sjunnesson1992cars}.
The computational domain is relatively simple and inexpensive to simulate, though, 
and it is a well-studied test case for non-swirling turbulent 
flames~\cite{volvo1,volvo2,volvo3,volvo4,COCKS20153394,ROCHETTE2018417}.
Furthermore, experimental velocity, turbulence statistics, and temperature measurements are 
available for two inlet velocities and temperatures for validation 
purposes~\cite{sjunnesson1991lda,sjunnesson1991validation,sjunnesson1992cars}.
Section S2 of the Supplemental Material contains a validation of our setup
using the non-reacting simulated velocity and turbulence 
statistics, compared with available experimental data; we did not include this 
in the main text, since our main focus is computational performance of the reacting solver.

\begin{figure}[htbp]
 \centering
 \begin{subfigure}{0.8\textwidth}
 \includegraphics[width=\linewidth]{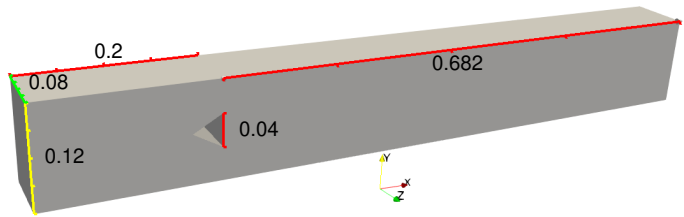}
 \caption{Schematic of the computational domain for the Volvo bluff-body stabilized flame case.
	  All distances are in meters.}
 \label{F:bluffbody_schematic}
 \end{subfigure}\\%
 \begin{subfigure}{0.8\textwidth}
 \includegraphics[width=\linewidth]{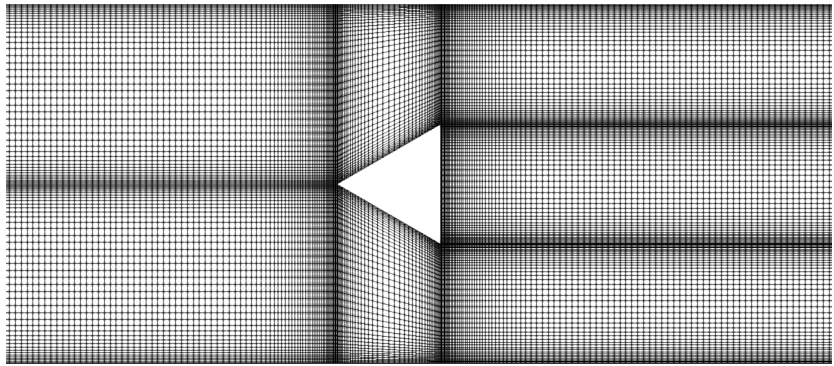}
 \caption{A close up of the computational mesh near the bluff body.
	  The cells are gradually stretched from a maximum wall-normal distance of \SI{0.3}{\milli\meter} near the walls and bluff body to a nominal mesh size of \SI{2}{\milli\meter}.}
 \label{F:bluffbody_mesh}
 \end{subfigure}
 \caption{The computational (a) domain and (b) mesh for the Volvo bluff-body stabilized flame case.}
 \label{F:bluffbody}
\end{figure}

Figure~\ref{F:bluffbody_schematic} shows the computational domain we used here, which
omits the fuel injection, seeding, and flow-straightener parts upstream of the bluff body. 
Instead, we set steady inflow boundary conditions at $\num{5}D$---where $D=\SI{40}{\milli\meter}$ 
is the bluff-body height---upstream of the trailing edge of the bluff body, as used in previous 
studies~\cite{volvo2,mvp,COCKS20153394}. The domain extends \SI{0.662}{\meter} downstream 
of the trailing edge of the bluff body, where we used wave-transmissive outflow conditions
as suggested in previous work~\cite{mvp,volvo2,volvo4}.
The domain is $\num{2}D$ wide in the span-wise direction; the front and back faces of the domain 
use periodic boundary conditions to reduce computational effort~\cite{mvp,COCKS20153394}.
The walls of the domain use adiabatic and no-slip boundary conditions.
\cref{T:bluffbody_conditions} lists the inlet conditions
, based on the available experimental conditions~\cite{mvp_exp}.

We developed the computational mesh (shown in \cref{F:bluffbody_mesh}) based on recommendations
from previous studies~\cite{COCKS20153394,mvp}; it uses hexagonal cells nominally 
\SI{2}{\milli\meter} large and a maximum wall-normal distance of \SI{0.3}{\milli\meter}.
Grading clusters the cells near the walls and bluff-body edges, as well as the shear 
layers downstream of the bluff body; the domain is as isotropic as possible elsewhere, 
resulting in a total of \num{2365000} mesh cells.

\begin{table}[htbp]
 \centering
 \sisetup{table-format=1.2}
 \begin{tabular}{@{}c c c@{}}
 \toprule
  Parameter 	 & Non-reactive & Reactive \\
  \midrule
  Pressure       & \SI{1}{\bar} 		    & \SI{1}{\bar} \\
  Temperature	 & \SI{288}{\kelvin}	   	    & \SI{288}{\kelvin} \\
  $U_\text{bulk}$& \SI{16.6}{\meter\per\second}	    & \SI{17.3}{\meter\per\second} \\
  \bottomrule
 \end{tabular}
 \caption{Inlet conditions for non-reactive and reactive cases~\cite{mvp}, selected to match 
 the available archived experiment data~\cite{mvp_exp}.
 $U_\text{bulk}$ is the bulk inlet velocity.}
 \label{T:bluffbody_conditions}
\end{table}

To simulate this case we used the \texttt{OpenFOAM} solver \texttt{reactingFoam} with the LES turbulence model, 
along with a Smagorinsky subgrid scale model~\cite{smag}. A second-order accurate 
central-differencing scheme discretized the Laplacian terms, while the solver used a second-order 
bounded cell-based Green-gauss method for gradient calculations.
We used a second-order weakly-limited central scheme to calculate the velocity advection 
divergence and strong limiting for turbulent kinetic energy, species, and energy scalars.
The solver advanced in time with a blended first- and second-order Crank--Nicolson scheme, and 
we handled pressure-velocity coupling with the PISO (pressure-implicit with split operator) 
algorithm~\cite{Issa1986}. Using a fully implicit second-order time-stepping method 
(along with an unlimited velocity advection scheme) was possible with multiple outer corrector 
steps---i.e., the PIMPLE algorithm in place of PISO---to avoid breakup of the flow due to 
pressure-velocity decoupling. However, this would require multiple solutions of the chemical 
kinetic ODEs at each step of the simulation, and consequentially make completing a single 
time step using the \texttt{OpenFOAM} ODE solvers take longer than the largest available time reservation 
on the computing cluster we used here. Hence, we used the blended time stepping and weakly 
limited velocity advection schemes instead.

\subsubsection{Results}
\label{S:bluffbody_reactive}

We ran the performance studies on 96 cores (i.e., four nodes with 2$\times$12 cores on each) 
of an Intel E5-2690 V3 CPU, with \texttt{AVX2} vector instructions, \SI{128}{\giga\byte} of RAM, and 
using the Intel OpenCL runtime \texttt{v16.1.1}.
We used \texttt{OpenFOAM} version \texttt{6.x}~\cite{OFV6} with \texttt{v2.1.0} of the OpenMPI library~\cite{open_mpi}, 
compiled with \texttt{gcc v5.4.0}~\cite{Stallman:2009:UGC:1593499}.
We instrumented the \texttt{reactingFoam} solver with the MPI-profiling library 
\texttt{IPM v2.0.6}~\cite{skinner2005performance} by placing profiling sections (via~\texttt{MPI\_PControl} calls) 
around the calls to the turbulent combustion model, ODE integration, and other key parts of the CFD time step 
(e.g., convection evaluation). The absolute and relative tolerances of the \texttt{OpenFOAM} and \texttt{accelerInt} \texttt{ROS4} solvers 
were \num{e-10} and \num{e-6}, respectively.

Although the Volvo bluff-body reactive flame experiments used a lean propane-air mixture (overall global equivalence 
ratio of $\phi=\num{0.6}$), here we chose a stoichiometric \ce{CH4}\slash air combustion case for this study.
Previous studies have shown mixed agreement between simulated and experimental results using \texttt{OpenFOAM} on this 
configuration~\cite{COCKS20153394,volvo4}. Considering the many potential sources of error in a reactive-flow 
simulation~\cite{ROCHETTE2018417}---e.g., the quality of chemical kinetic model, the turbulence-chemistry 
interaction model, and the numerical solver itself~\cite{COCKS20153394}---we opted to focus on the performance 
of the ODE integration algorithms instead of a detailed comparison of the reactive-flow case to the experimental 
data (as done for the non-reactive case as shown in the Supplemental Material).
While we expect that these results could be reproduced for, e.g., the UCSD propane model~\cite{ucsd} with 
\num{57} species and \num{268} reactions, we prefer to simply use the comparably sized 
GRI-Mech 3.0~\cite{smith_gri-mech_30} for our demonstration.

We initialized the reactive-flow case using a RANS simulation with the coarse (\SI{4}{\milli\meter} nominal 
cell size) mesh and the base \texttt{OpenFOAM} \texttt{ROS4} integrator. A two-step methane model~\cite{FRANZELLI2012621} ignited 
the flame via a spherical high-temperature kernel placed behind the bluff body.
After the flame stabilized and attached to the bluff body, we ran the RANS simulation using the two-step 
methane model for a single flow-through time to develop the temperature field. Next, we mapped the RANS solution
onto the same coarse mesh using the full GRI-Mech 3.0 model with the numerical schemes and LES setup described above;
we reduced the CFD time-step size to \SI{3e-7}{\second} such that the maximum Courant number remained 
under \num{0.05}, and set a minimum reacting temperature threshold of \SI{550}{\kelvin}.
As species transport is more relevant in the reactive case, we used the Sutherland transport model in \texttt{OpenFOAM}, 
and determined species coefficients by a non-linear least squares fit~\cite{scipy,Virtanen2020} to the species 
viscosities versus temperature relation obtained from \texttt{Cantera}~\cite{cantera}.

\begin{figure}[htbp]
 \includegraphics[width=\textwidth]{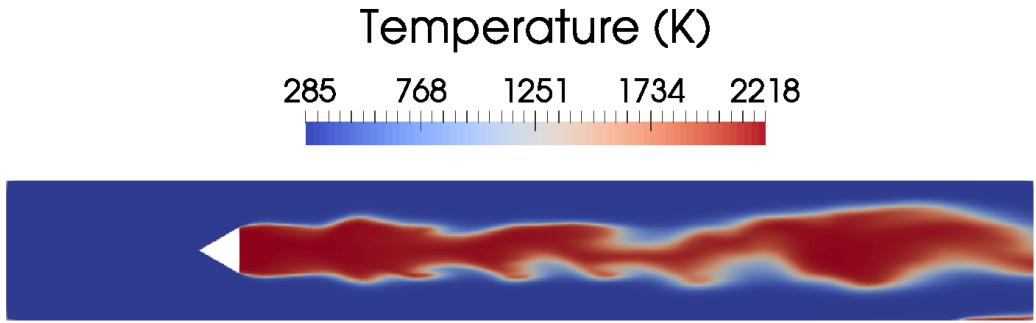}
 \caption{Instantaneous temperature contour of the reactive case on the finer LES grid.}
 \label{F:temperature_contour}
\end{figure}

For the reactive case, the \texttt{OpenFOAM} \texttt{ROS4} solver proved too slow to advance the solution in a reasonable 
time frame, so the \texttt{accelerInt} solver was used instead to run the coarse LES mesh for a full flow-through 
time to develop the solution. Next, the coarse solution was mapped onto the finer grid described 
in~\cref{sbb:case} and run for \SI{3}{\milli\second} of simulated time using the \texttt{accelerInt} \texttt{ROS4} solver
to obtain baseline timing and solution data. \Cref{F:temperature_contour} shows an instantaneous snapshot 
of the temperature field on the finer LES grid. For the fine mesh however, the \texttt{OpenFOAM} \texttt{ROS4} solver was 
incapable of finishing more than a single CFD time step in the maximum allotted time reservation 
(\SI{6}{\hour}) on the computing cluster used here, with each step taking on average, 
\SIrange{3.02}{3.80}{\hour} of wall-clock time\footnote{Here, the mean wall-clock 
time per CFD step is normalized by the total number of MPI ranks in the simulation such that it can be 
directly compared to the maximum time-reservation on the cluster. The mean wall-clock times per CFD step 
later reported in~\cref{T:performance} do not have this normalization applied.}; 
in contrast, we note that a single time step using the \texttt{accelerInt} solver took only an average of 
\SI{6.5}{\minute} of wall-clock time. Thus, simulating the entire \SI{3}{\milli\second} duration 
with the \texttt{OpenFOAM} solver would have required the completion of over \num{10000} individual job submissions.
To get around this issue, \num{150} evenly spaced solution points were selected throughout the 
\SI{3}{\milli\second} duration, from each of which the solution was computed for a single CFD time step 
using the \texttt{OpenFOAM} \texttt{ROS4} solver for comparison with \texttt{accelerInt}.

The IPM library provides the total wall-clock execution time spent in user-designated MPI profiling 
sections for each MPI rank in the simulation over a single job run.
Therefore, the computational time spent in, e.g., solving the chemical kinetic ODEs for each individual 
CFD time step cannot be obtained directly.
Instead, we normalized the total wall-clock execution time by the number of CFD time steps completed in a run, 
and averaged the normalized wall-clock time per CFD time step over the total number of runs 
(\num{150} for the \texttt{OpenFOAM} solver, and \num{175} for \texttt{accelerInt}).

\begin{table}[htbp]
 \centering
 \captionsetup{width=0.8\textwidth}
 \sisetup{table-format=2.2,separate-uncertainty=true,multi-part-units=single}
 \begin{tabular}{@{}l l S S@{}}
 \toprule
 Region & Quantity & \multicolumn{1}{c}{\texttt{accelerInt}} & \multicolumn{1}{c}{\texttt{OpenFOAM}} \\
 \midrule
 \multirow{3}{*}{ODE solution}   & Time 	& \SI{1.90\pm.192}{\hour} 	& \SI{58.95\pm8.05}{\hour} \\
			         & Speedup 	& $\SI{31.06\pm5.27}{\times}$ 	& \multicolumn{1}{c}{--} \\
			         & \% of total	& $\SI{18.79\pm3.32}{\percent}$ & $\SI{20.86\pm2.93}{\percent}$ \\
 \midrule
 \multirow{3}{*}{Load balancing} & Time		& \SI{7.11\pm0.77}{\hour} 	& \SI{254.34\pm28.60}{\hour} \\
			         & Speedup 	& $\SI{35.82\pm5.59}{\times}$ 	& \multicolumn{1}{c}{--} \\
			         & \% of total	& $\SI{81.06\pm12.89}{\percent}$ & $\SI{78.06\pm11.64}{\percent}$ \\
 \midrule
 \multirow{3}{*}{Total} 	 & Time 	& \SI{9.10\pm0.94}{\hour} 	& \SI{313.76\pm35.28}{\hour} \\
			         & Speedup 	& $\SI{34.48\pm5.25}{\times}$ 	& \multicolumn{1}{c}{--} \\
			         & \% of total	& \SI{100}{$\percent$} 		& \SI{100}{$\percent$} \\
 \bottomrule
 \end{tabular}
 \caption{The mean wall-clock execution time per CFD time step, speedup, and percent of total 
 runtime spent in ODE integration and load balancing for both the \texttt{accelerInt} and \texttt{OpenFOAM} \texttt{ROS4} ODE 
 integration solvers averaged over ten runs. The times are reported in CPU hours per step, 
 and were run on \num{96} Intel E5-2690 V3 CPU cores with \texttt{AVX2} vector instructions.}
 \label{T:performance}
\end{table}

\cref{T:performance} presents the average wall-clock execution time per CFD time step for the 
\texttt{accelerInt} and \texttt{OpenFOAM} solvers. The \texttt{accelerInt} \texttt{ROS4} solver performs, on average, \num{32.65}$\times$
faster (with a variation of \num{\pm1.87}$\times$ depending on the run) in solving the chemical 
kinetic ODEs than the corresponding \texttt{OpenFOAM} approach. In addition, \texttt{OpenFOAM} poorly balances the load of 
the chemistry integration, with both the \texttt{accelerInt} and \texttt{OpenFOAM} chemistry solvers spending upwards of 
\SI{75}{$\percent$} of the total computational time waiting on the solution of the chemical kinetic 
ODEs from other MPI ranks. \texttt{OpenFOAM} uses a simple static decomposition of the domain to distribute the 
computational cells to the various MPI ranks; hence, if a majority of the cells on an MPI rank are 
considered ``non-reactive'' (i.e., the temperature of the cell is less than the \SI{550}{\kelvin} 
threshold) the CPU core will sit idle during chemistry integration.
A stiffness-based chemical kinetic load balancing scheme (e.g., as implemented by 
Kodavasal et al.~\cite{kodavasal2016development}) could further improve the 
computational efficiency of reactive-flow simulations in \texttt{OpenFOAM}, and merits future investigation.
Finally,~\cref{T:performance} shows that the bluff-body simulation in \texttt{OpenFOAM} using the \texttt{accelerInt} solver 
spends \num{35.82\pm5.59}$\times$ less time waiting due to poor chemical kinetic load balancing, 
while the overall simulation is \num{34.48\pm5.25}$\times$ faster.

\section{Lessons learned}

Our strategy for vectorizing chemistry integration focused on building a solver based on extended
versions of existing models in \texttt{OpenFOAM}. This consisted of
\begin{itemize}
  \item existing turbulence-combustion interaction models (e.g., EDC, partially stirred reactor [PaSR]) to enable
  profiling via IPM~\cite{skinner2005performance};
  \item a batched chemistry model, based on the \texttt{StandardChemistryModel} class in \texttt{OpenFOAM} that additionally
  enables optional IPM profiling, that (1) scans the list of cells to determine the list of interacting cells based 
  on the temperature cut-off, (2) converts the state variables to the form expected by \texttt{pyJac}, 
  (3) passes the complete list of reacting cells to the \texttt{accelerInt} solver(s), and 
  (4) finally converts state variables for the integration back to the native \texttt{OpenFOAM} format and stores 
  them in the appropriate buffers;
  \item three solvers (each with optional IPM profiling):
  \begin{itemize}
    \item \texttt{testODE}, used to verify the accuracy of the \texttt{accelerInt} solvers when converting to and 
    from the \texttt{OpenFOAM} state variables (as compared to Cantera),
    \item \texttt{chemFoam}, used to verify the \texttt{accelerInt} solver(s) coupling to \texttt{OpenFOAM} (i.e., integrating the 
    system of equations), and
    \item \texttt{reactingFoam}, used to enforce the explicit mass conservation in \texttt{pyJac}'s state variables, 
    and enable the coupling to \texttt{accelerInt} for production simulations.
  \end{itemize}
\end{itemize}
This strategy allowed us to vectorize chemistry integration while modifying little of \texttt{OpenFOAM} itself.
Instead, almost all the logic for enabling alternate models and solvers was contained in the 
\texttt{reactingFoamIPM} module.
Simultaneously, because we built the \texttt{reactingFoamIPM} module against existing \texttt{OpenFOAM} classes, 
we had to implement few new features there, instead using calls to the parent class wherever possible.
For example, a call to the IPM-profiling-enabled PaSR class's correction method just required placing 
\texttt{MPI\_Pcontrol} calls around a call to the base PaSR class's correction method.
Thus, we did not need to vectorize \texttt{OpenFOAM} itself---the OpenCL code was entirely contained within the 
\texttt{accelerInt} and \texttt{pyJac} libraries.

We pursued an OpenCL-based strategy here to vectorize code, with the intent of supporting 
portability to different platforms. However, in practice, we found that this led to frequent 
problems involving segmentation faults during compiling, program crashes, incorrect compiled code, 
and more, in addition to the associated increase in complexity of the source code.
Ultimately, this limited the utility of our approach.
Moving forward, we recommend examining a pure OpenMP-based implementation to enable broad adoption;
OpenMP version 5.0~\cite{dagum1998openmp,openmp5} might also support offloading calculations to GPUs.

\section{Conclusions}
\label{sbb:conc}

This work adapted several previously developed linearly implicit and explicit OpenCL-vectorized 
ODE integration methods~\cite{stone2018} into the \texttt{accelerInt} library~\cite{accelerInt:beta2}.
We verified the accuracy of these solvers for the solution of the chemical kinetic ODEs (using source-rate and 
analytical Jacobian evaluation codes from the \texttt{pyJac} code-generation platform~\cite{CURTIS2018186}) 
for a variety of cases in this effort.
We developed \texttt{OpenFOAM}-based models of the Sandia Flame D~\cite{BARLOW19981087,BARLOW2005433,SCHNEIDER2003185} 
stabilized jet flame and Volvo bluff-body stabilized premixed flame~\cite{bluffbody_les_case}.
Furthermore, we used a vectorized solver to accelerate chemical-kinetic integration of these reactive simulations,
resulting in large performance improvements.

The major contributions of this work are
\begin{itemize}
 \item Incorporating vectorized chemical kinetic ODE integration methods (provided by the \texttt{accelerInt} and \texttt{pyJac} 
 libraries) into a publicly available \texttt{OpenFOAM} solver~\cite{reactingfoamipm};

 \item Achieving significant speedups (\numrange{33}{35}$\times$) via the vectorized solvers over built-in \texttt{OpenFOAM} 
 integration methods on the same hardware (an Intel Xeon CPU with AVX2 vector instructions); and

 \item Profiling the solution of realistic reactive-flow simulations in \texttt{OpenFOAM}, demonstrating the need 
 for chemical kinetic load-balancing.
\end{itemize}

Future extensions to this work should focus on a few key aspects.
First, this study has demonstrated that one of the largest bottlenecks of realistic reactive-flow 
simulations in \texttt{OpenFOAM} is the poor chemical kinetic load balancing based on static decomposition.
Over \SI{75}{$\percent$} of the overall computational time is wasted using either the \texttt{accelerInt} or native 
\texttt{OpenFOAM} solvers due to this issue. A number of chemical kinetic load balancing algorithms 
exist~\cite{Flowers2006,Shi2009}, but the stiffness-based load-balancing technique of 
Kodavasal et al.~\cite{kodavasal2016development} seems particularly promising.

More generally, this work has shown that vectorized, linearly implicit ODE integration 
(paired with analytical Jacobian evaluation) can greatly accelerate reactive-flow simulations.
Developing vectorized sparse linear algebra\slash matrix factorization codes would be an 
excellent---while challenging---extension of this work, further speeding up the integration of 
larger detailed chemical models. More advanced integration algorithms should also be investigated; 
W-methods~\cite{wanner1991solving} are particularly promising for their ability to re-use previously
evaluated Jacobians and LU-factorizations without needing Newton iteration.

We also found challenges in the OpenCL-based programming approach, which we originally 
pursued for portability. Moving forward, we recommend using pure OpenMP for vectorization,
which should avoid substantially increasing source complexity and also may enable 
offloading of calculations to GPUs.

\section*{Acknowledgments}
This material is based upon work supported by the National Science Foundation under grants 
ACI-1534688 (Curtis and Sung) and ACI-1535065 (Niemeyer). We would also like to thank Christopher Stone of 
Computational Science and Engineering, LLC for making available the OpenCL integrators used in this study.

\appendix
\setcounter{figure}{0}
\setcounter{table}{0}
\renewcommand*{\thesection}{\appendixname~\Alph{section}}

\section*{Appendix: Availability of material}
\label{A:availability}
The vectorized ODE integration methods used in this work are available in version 
\texttt{2.0-beta} of the \texttt{accelerInt} library~\cite{accelerInt:beta2}, while 
\texttt{pyJac}-\texttt{v2.0}~\cite{pyJac_v2} provides the chemical kinetic source terms and Jacobian 
evaluation. The \texttt{OpenFOAM} case modeling the Volvo bluff-body flame is available for download 
via GitHub~\cite{bluffbody_les_case}, and the coupled \texttt{OpenFOAM}\slash\texttt{accelerInt} solver 
(with \texttt{IPM}-based profiling) is also made available for public use~\cite{reactingfoamipm}.
Finally, the data, figures, and plotting scripts used to generate this paper are 
similarly available for public use~\cite{paper_data}.

\bibliography{bluffbody_les}{}

\begin{thebibliography}{110}
\expandafter\ifx\csname natexlab\endcsname\relax\def\natexlab#1{#1}\fi
\providecommand{\url}[1]{\texttt{#1}}
\providecommand{\href}[2]{#2}
\providecommand{\path}[1]{#1}
\providecommand{\DOIprefix}{doi:}
\providecommand{\ArXivprefix}{arXiv:}
\providecommand{\URLprefix}{URL: }
\providecommand{\Pubmedprefix}{pmid:}
\providecommand{\doi}[1]{\href{http://dx.doi.org/#1}{\path{#1}}}
\providecommand{\Pubmed}[1]{\href{pmid:#1}{\path{#1}}}
\providecommand{\bibinfo}[2]{#2}
\ifx\xfnm\relax \def\xfnm[#1]{\unskip,\space#1}\fi
\bibitem[{Imtenan et~al.(2014)Imtenan, Varman, Masjuki, Kalam, Sajjad, Arbab,
  and Fattah}]{IMTENAN2014329}
\bibinfo{author}{S.~Imtenan}, \bibinfo{author}{M.~Varman},
  \bibinfo{author}{H.~Masjuki}, \bibinfo{author}{M.~Kalam},
  \bibinfo{author}{H.~Sajjad}, \bibinfo{author}{M.~Arbab},
  \bibinfo{author}{I.~R. Fattah},
\newblock \bibinfo{title}{Impact of low temperature combustion attaining
  strategies on diesel engine emissions for diesel and biodiesels: A review},
\newblock \bibinfo{journal}{Energy Convers. Manag.} \bibinfo{volume}{80}
  (\bibinfo{year}{2014}) \bibinfo{pages}{329--356}.
  \DOIprefix\doi{10.1016/j.enconman.2014.01.020}.
\bibitem[{Westbrook et~al.(2005)Westbrook, Mizobuchi, Poinsot, Smith, and
  Warnatz}]{WESTBROOK2005125}
\bibinfo{author}{C.~K. Westbrook}, \bibinfo{author}{Y.~Mizobuchi},
  \bibinfo{author}{T.~J. Poinsot}, \bibinfo{author}{P.~J. Smith},
  \bibinfo{author}{J.~Warnatz},
\newblock \bibinfo{title}{Computational combustion},
\newblock \bibinfo{journal}{Proc. Combust. Inst.} \bibinfo{volume}{30}
  (\bibinfo{year}{2005}) \bibinfo{pages}{125--157}.
  \DOIprefix\doi{10.1016/j.proci.2004.08.275}.
\bibitem[{Moiz et~al.(2016)Moiz, Ameen, Lee, and Som}]{Moiz2016123}
\bibinfo{author}{A.~A. Moiz}, \bibinfo{author}{M.~M. Ameen},
  \bibinfo{author}{S.-Y. Lee}, \bibinfo{author}{S.~Som},
\newblock \bibinfo{title}{Study of soot production for double injections of
  n-dodecane in {CI} engine-like conditions},
\newblock \bibinfo{journal}{Combust. Flame} \bibinfo{volume}{173}
  (\bibinfo{year}{2016}) \bibinfo{pages}{123--131}.
  \DOIprefix\doi{10.1016/j.combustflame.2016.08.005}.
\bibitem[{Lu and Law(2009)}]{Lu:2009gh}
\bibinfo{author}{T.~Lu}, \bibinfo{author}{C.~K. Law},
\newblock \bibinfo{title}{Toward accommodating realistic fuel chemistry in
  large-scale computations},
\newblock \bibinfo{journal}{Prog. Energy Comb. Sci.} \bibinfo{volume}{35}
  (\bibinfo{year}{2009}) \bibinfo{pages}{192--215}.
  \DOIprefix\doi{10.1016/j.pecs.2008.10.002}.
\bibitem[{Tur{\'a}nyi and Tomlin(2014)}]{Turanyi:2014aa}
\bibinfo{author}{T.~Tur{\'a}nyi}, \bibinfo{author}{A.~S. Tomlin},
  \bibinfo{title}{Analysis of Kinetic Reaction Mechanisms},
  \bibinfo{publisher}{Springer-Verlag}, \bibinfo{address}{Berlin Heidelberg},
  \bibinfo{year}{2014}. \DOIprefix\doi{10.1007/978-3-662-44562-4}.
\bibitem[{Lu and Law(2006)}]{Lu:2006bb}
\bibinfo{author}{T.~Lu}, \bibinfo{author}{C.~K. Law},
\newblock \bibinfo{title}{Linear time reduction of large kinetic mechanisms
  with directed relation graph: \emph{n}-heptane and iso-octane},
\newblock \bibinfo{journal}{Combust. Flame} \bibinfo{volume}{144}
  (\bibinfo{year}{2006}) \bibinfo{pages}{24--36}.
  \DOIprefix\doi{10.1016/j.combustflame.2005.02.015}.
\bibitem[{Pepiot-Desjardins and Pitsch(2008)}]{Pepiot-Desjardins:2008}
\bibinfo{author}{P.~Pepiot-Desjardins}, \bibinfo{author}{H.~Pitsch},
\newblock \bibinfo{title}{An efficient error-propagation-based reduction method
  for large chemical kinetic mechanisms},
\newblock \bibinfo{journal}{Combust. Flame} \bibinfo{volume}{154}
  (\bibinfo{year}{2008}) \bibinfo{pages}{67--81}.
  \DOIprefix\doi{10.1016/j.combustflame.2007.10.020}.
\bibitem[{Hiremath et~al.(2010)Hiremath, Ren, and Pope}]{Hiremath:2010jw}
\bibinfo{author}{V.~Hiremath}, \bibinfo{author}{Z.~Ren}, \bibinfo{author}{S.~B.
  Pope},
\newblock \bibinfo{title}{A greedy algorithm for species selection in dimension
  reduction of combustion chemistry},
\newblock \bibinfo{journal}{Combust. Theor. Model.} \bibinfo{volume}{14}
  (\bibinfo{year}{2010}) \bibinfo{pages}{619--652}.
  \DOIprefix\doi{10.1080/13647830.2010.499964}.
\bibitem[{Niemeyer et~al.(2010)Niemeyer, Sung, and Raju}]{Niemeyer:2010bt}
\bibinfo{author}{K.~E. Niemeyer}, \bibinfo{author}{C.~J. Sung},
  \bibinfo{author}{M.~P. Raju},
\newblock \bibinfo{title}{Skeletal mechanism generation for surrogate fuels
  using directed relation graph with error propagation and sensitivity
  analysis},
\newblock \bibinfo{journal}{Combust. Flame} \bibinfo{volume}{157}
  (\bibinfo{year}{2010}) \bibinfo{pages}{1760--1770}.
  \DOIprefix\doi{10.1016/j.combustflame.2009.12.022}.
\bibitem[{Lu and Law(2007)}]{Lu:2007}
\bibinfo{author}{T.~Lu}, \bibinfo{author}{C.~K. Law},
\newblock \bibinfo{title}{Diffusion coefficient reduction through species
  bundling},
\newblock \bibinfo{journal}{Combust. Flame} \bibinfo{volume}{148}
  (\bibinfo{year}{2007}) \bibinfo{pages}{117--126}.
  \DOIprefix\doi{10.1016/j.combustflame.2006.10.004}.
\bibitem[{Ahmed et~al.(2007)Ahmed, Mau{\ss}, Mor{\'e}ac, and
  Zeuch}]{Ahmed:2007fa}
\bibinfo{author}{S.~S. Ahmed}, \bibinfo{author}{F.~Mau{\ss}},
  \bibinfo{author}{G.~Mor{\'e}ac}, \bibinfo{author}{T.~Zeuch},
\newblock \bibinfo{title}{A comprehensive and compact \emph{n}-heptane
  oxidation model derived using chemical lumping},
\newblock \bibinfo{journal}{Phys. Chem. Chem. Phys.} \bibinfo{volume}{9}
  (\bibinfo{year}{2007}) \bibinfo{pages}{1107--1126}.
  \DOIprefix\doi{10.1039/b614712g}.
\bibitem[{Pepiot-Desjardins and Pitsch(2008)}]{Pepiot:2008kq}
\bibinfo{author}{P.~Pepiot-Desjardins}, \bibinfo{author}{H.~Pitsch},
\newblock \bibinfo{title}{An automatic chemical lumping method for the
  reduction of large chemical kinetic mechanisms},
\newblock \bibinfo{journal}{Combust. Theor. Model.} \bibinfo{volume}{12}
  (\bibinfo{year}{2008}) \bibinfo{pages}{1089--1108}.
  \DOIprefix\doi{10.1080/13647830802245177}.
\bibitem[{Maas and Pope(1992)}]{Maas:1992aa}
\bibinfo{author}{U.~Maas}, \bibinfo{author}{S.~B. Pope},
\newblock \bibinfo{title}{Simplifying chemical kinetics: intrinsic
  low-dimensional manifolds in composition space},
\newblock \bibinfo{journal}{Combust. Flame} \bibinfo{volume}{88}
  (\bibinfo{year}{1992}) \bibinfo{pages}{239--264}.
  \DOIprefix\doi{10.1016/0010-2180(92)90034-M}.
\bibitem[{Lam and Goussis(1994)}]{Lam:1994ws}
\bibinfo{author}{S.-H. Lam}, \bibinfo{author}{D.~A. Goussis},
\newblock \bibinfo{title}{The {CSP} method for simplying kinetics},
\newblock \bibinfo{journal}{Int. J. Chem. Kinet.} \bibinfo{volume}{26}
  (\bibinfo{year}{1994}) \bibinfo{pages}{461--486}.
  \DOIprefix\doi{10.1002/kin.550260408}.
\bibitem[{Lu et~al.(2001)Lu, Ju, and Law}]{Lu:2001ve}
\bibinfo{author}{T.~Lu}, \bibinfo{author}{Y.~Ju}, \bibinfo{author}{C.~K. Law},
\newblock \bibinfo{title}{Complex {CSP} for chemistry reduction and analysis},
\newblock \bibinfo{journal}{Combust. Flame} \bibinfo{volume}{126}
  (\bibinfo{year}{2001}) \bibinfo{pages}{1445--1455}.
  \DOIprefix\doi{10.1016/S0010-2180(01)00252-8}.
\bibitem[{Gou et~al.(2010)Gou, Sun, Chen, and Ju}]{Gou:2010}
\bibinfo{author}{X.~Gou}, \bibinfo{author}{W.~Sun}, \bibinfo{author}{Z.~Chen},
  \bibinfo{author}{Y.~Ju},
\newblock \bibinfo{title}{A dynamic multi-timescale method for combustion
  modeling with detailed and reduced chemical kinetic mechanisms},
\newblock \bibinfo{journal}{Combust. Flame} \bibinfo{volume}{157}
  (\bibinfo{year}{2010}) \bibinfo{pages}{1111--1121}.
  \DOIprefix\doi{10.1016/j.combustflame.2010.02.020}.
\bibitem[{Pope(1997)}]{Pope:1997wu}
\bibinfo{author}{S.~B. Pope},
\newblock \bibinfo{title}{Computationally efficient implementation of
  combustion chemistry using in situ adaptive tabulation},
\newblock \bibinfo{journal}{Combust. Theor. Model.} \bibinfo{volume}{1}
  (\bibinfo{year}{1997}) \bibinfo{pages}{41--63}.
  \DOIprefix\doi{10.1080/713665229}.
\bibitem[{Ren et~al.(2014)Ren, Liu, Lu, Lu, Oluwole, and Goldin}]{Ren:2014cd}
\bibinfo{author}{Z.~Ren}, \bibinfo{author}{Y.~Liu}, \bibinfo{author}{T.~Lu},
  \bibinfo{author}{L.~Lu}, \bibinfo{author}{O.~O. Oluwole},
  \bibinfo{author}{G.~M. Goldin},
\newblock \bibinfo{title}{The use of dynamic adaptive chemistry and tabulation
  in reactive flow simulations},
\newblock \bibinfo{journal}{Combust. Flame} \bibinfo{volume}{161}
  (\bibinfo{year}{2014}) \bibinfo{pages}{127--137}.
  \DOIprefix\doi{10.1016/j.combustflame.2013.08.018}.
\bibitem[{Li et~al.(2018)Li, Lewandowski, Contino, and Parente}]{tdac}
\bibinfo{author}{Z.~Li}, \bibinfo{author}{M.~T. Lewandowski},
  \bibinfo{author}{F.~Contino}, \bibinfo{author}{A.~Parente},
\newblock \bibinfo{title}{Assessment of on-the-fly chemistry reduction and
  tabulation approaches for the simulation of moderate or intense low-oxygen
  dilution combustion},
\newblock \bibinfo{journal}{Energy Fuels} \bibinfo{volume}{32}
  (\bibinfo{year}{2018}) \bibinfo{pages}{10121--10131}.
  \DOIprefix\doi{10.1021/acs.energyfuels.8b01001}.
\bibitem[{Lu and Law(2008)}]{Lu:2008bi}
\bibinfo{author}{T.~Lu}, \bibinfo{author}{C.~K. Law},
\newblock \bibinfo{title}{Strategies for mechanism reduction for large
  hydrocarbons: n-heptane},
\newblock \bibinfo{journal}{Combust. Flame} \bibinfo{volume}{154}
  (\bibinfo{year}{2008}) \bibinfo{pages}{153--163}.
  \DOIprefix\doi{10.1016/j.combustflame.2007.11.013}.
\bibitem[{Niemeyer and Sung(2014)}]{Niemeyer:2014}
\bibinfo{author}{K.~E. Niemeyer}, \bibinfo{author}{C.~J. Sung},
\newblock \bibinfo{title}{Mechanism reduction for multicomponent surrogates: A
  case study using toluene reference fuels},
\newblock \bibinfo{journal}{Combust. Flame} \bibinfo{volume}{161}
  (\bibinfo{year}{2014}) \bibinfo{pages}{2752--2764}.
  \DOIprefix\doi{10.1016/j.combustflame.2014.05.001}.
\bibitem[{Niemeyer and Sung(2015)}]{Niemeyer:2015wq}
\bibinfo{author}{K.~E. Niemeyer}, \bibinfo{author}{C.~J. Sung},
\newblock \bibinfo{title}{Reduced chemistry for a gasoline surrogate valid at
  engine-relevant conditions},
\newblock \bibinfo{journal}{Energy Fuels} \bibinfo{volume}{29}
  (\bibinfo{year}{2015}) \bibinfo{pages}{1172--1185}.
  \DOIprefix\doi{10.1021/ef5022126}.
\bibitem[{Liang et~al.(2009)Liang, Stevens, and Farrell}]{Liang:2009}
\bibinfo{author}{L.~Liang}, \bibinfo{author}{J.~Stevens},
  \bibinfo{author}{J.~T. Farrell},
\newblock \bibinfo{title}{A dynamic adaptive chemistry scheme for reactive flow
  computations},
\newblock \bibinfo{journal}{Proc. Combust. Inst.} \bibinfo{volume}{32}
  (\bibinfo{year}{2009}) \bibinfo{pages}{527--534}.
  \DOIprefix\doi{10.1016/j.proci.2008.05.073}.
\bibitem[{Yang et~al.(2013)Yang, Ren, Lu, and Goldin}]{Yang:2013ip}
\bibinfo{author}{H.~Yang}, \bibinfo{author}{Z.~Ren}, \bibinfo{author}{T.~Lu},
  \bibinfo{author}{G.~M. Goldin},
\newblock \bibinfo{title}{Dynamic adaptive chemistry for turbulent flame
  simulations},
\newblock \bibinfo{journal}{Combust. Theor. Model.} \bibinfo{volume}{17}
  (\bibinfo{year}{2013}) \bibinfo{pages}{167--183}.
  \DOIprefix\doi{10.1080/13647830.2012.733825}.
\bibitem[{Curtis et~al.(2015)Curtis, Niemeyer, and Sung}]{Curtis:2015}
\bibinfo{author}{N.~J. Curtis}, \bibinfo{author}{K.~E. Niemeyer},
  \bibinfo{author}{C.~J. Sung},
\newblock \bibinfo{title}{An automated target species selection method for
  dynamic adaptive chemistry simulations},
\newblock \bibinfo{journal}{Combust. Flame} \bibinfo{volume}{162}
  (\bibinfo{year}{2015}) \bibinfo{pages}{1358--1374}.
  \DOIprefix\doi{10.1016/j.combustflame.2014.11.004}.
\bibitem[{Mott et~al.(2000)Mott, Oran, and van Leer}]{Mott2000}
\bibinfo{author}{D.~R. Mott}, \bibinfo{author}{E.~S. Oran},
  \bibinfo{author}{B.~van Leer},
\newblock \bibinfo{title}{A quasi-steady-state solver for the stiff ordinary
  differential equations of reaction kinetics},
\newblock \bibinfo{journal}{Journal of Computational Physics}
  \bibinfo{volume}{164} (\bibinfo{year}{2000}) \bibinfo{pages}{407--428}.
  \DOIprefix\doi{10.1006/jcph.2000.6605}.
\bibitem[{Hansen and Sutherland(2017)}]{hansen18}
\bibinfo{author}{M.~A. Hansen}, \bibinfo{author}{J.~C. Sutherland},
\newblock \bibinfo{title}{Dual timestepping methods for detailed combustion
  chemistry},
\newblock \bibinfo{journal}{Combust. Theor. Model.} \bibinfo{volume}{21}
  (\bibinfo{year}{2017}) \bibinfo{pages}{329--345}.
  \DOIprefix\doi{10.1080/13647830.2016.1235728}.
\bibitem[{Shi et~al.(2012)Shi, Green, Wong, and Oluwole}]{Shi:2012aa}
\bibinfo{author}{Y.~Shi}, \bibinfo{author}{W.~H. Green}, \bibinfo{author}{H.-W.
  Wong}, \bibinfo{author}{O.~O. Oluwole},
\newblock \bibinfo{title}{Accelerating multi-dimensional combustion simulations
  using {GPU} and hybrid explicit\slash implicit {ODE} integration},
\newblock \bibinfo{journal}{Combust. Flame} \bibinfo{volume}{159}
  (\bibinfo{year}{2012}) \bibinfo{pages}{2388--2397}.
  \DOIprefix\doi{10.1016/j.combustflame.2012.02.016}.
\bibitem[{Stone and Bisetti(2014)}]{stone2014comparison}
\bibinfo{author}{C.~P. Stone}, \bibinfo{author}{F.~Bisetti},
\newblock \bibinfo{title}{Comparison of {ODE} solvers for chemical kinetics and
  reactive {CFD} applications},
\newblock in: \bibinfo{booktitle}{AIAA 52nd Aerospace Sciences Meeting
  (National Harbor, MD)}, \bibinfo{year}{2014}.
  \DOIprefix\doi{10.2514/6.2014-0822}, \bibinfo{note}{{AIAA} Paper No.
  2014-0822}.
\bibitem[{Niemeyer and Sung(2014)}]{Niemeyer:2014aa}
\bibinfo{author}{K.~E. Niemeyer}, \bibinfo{author}{C.~J. Sung},
\newblock \bibinfo{title}{Accelerating moderately stiff chemical kinetics in
  reactive-flow simulations using {GPUs}},
\newblock \bibinfo{journal}{J. Comput. Phys.} \bibinfo{volume}{256}
  (\bibinfo{year}{2014}) \bibinfo{pages}{854--871}.
  \DOIprefix\doi{10.1016/j.jcp.2013.09.025}.
\bibitem[{Imren and Haworth(2016)}]{IMREN20161}
\bibinfo{author}{A.~Imren}, \bibinfo{author}{D.~Haworth},
\newblock \bibinfo{title}{On the merits of extrapolation-based stiff {ODE}
  solvers for combustion {CFD}},
\newblock \bibinfo{journal}{Combust. Flame} \bibinfo{volume}{174}
  (\bibinfo{year}{2016}) \bibinfo{pages}{1--15}.
  \DOIprefix\doi{10.1016/j.combustflame.2016.09.018}.
\bibitem[{Curtis et~al.(2017)Curtis, Niemeyer, and Sung}]{CurtisGPU:2017}
\bibinfo{author}{N.~J. Curtis}, \bibinfo{author}{K.~E. Niemeyer},
  \bibinfo{author}{C.~J. Sung},
\newblock \bibinfo{title}{An investigation of {GPU}-based stiff chemical
  kinetics integration methods},
\newblock \bibinfo{journal}{Combust. Flame} \bibinfo{volume}{179}
  (\bibinfo{year}{2017}) \bibinfo{pages}{312--324}.
  \DOIprefix\doi{10.1016/j.combustflame.2017.02.005}.
\bibitem[{Stone et~al.(2018)Stone, Alferman, and Niemeyer}]{stone2018}
\bibinfo{author}{C.~P. Stone}, \bibinfo{author}{A.~T. Alferman},
  \bibinfo{author}{K.~E. Niemeyer},
\newblock \bibinfo{title}{Accelerating finite-rate chemical kinetics with
  coprocessors: comparing vectorization methods on {GPUs}, {MICs}, and {CPUs}},
\newblock \bibinfo{journal}{Comput. Phys. Comm.} \bibinfo{volume}{226}
  (\bibinfo{year}{2018}) \bibinfo{pages}{18--29}.
  \DOIprefix\doi{10.1016/j.cpc.2018.01.015}.
\bibitem[{Niemeyer et~al.(2017)Niemeyer, Curtis, and Sung}]{Niemeyer:2016aa}
\bibinfo{author}{K.~E. Niemeyer}, \bibinfo{author}{N.~J. Curtis},
  \bibinfo{author}{C.~J. Sung},
\newblock \bibinfo{title}{\texttt{pyJac}: analytical {Jacobian} generator for
  chemical kinetics},
\newblock \bibinfo{journal}{Comput. Phys. Comm.} \bibinfo{volume}{215}
  (\bibinfo{year}{2017}) \bibinfo{pages}{188--203}.
  \DOIprefix\doi{10.1016/j.cpc.2017.02.004}.
\bibitem[{Curtis et~al.(2018)Curtis, Niemeyer, and Sung}]{CURTIS2018186}
\bibinfo{author}{N.~J. Curtis}, \bibinfo{author}{K.~E. Niemeyer},
  \bibinfo{author}{C.~J. Sung},
\newblock \bibinfo{title}{Using {SIMD} and {SIMT} vectorization to evaluate
  sparse chemical kinetic {Jacobian} matrices and thermochemical source terms},
\newblock \bibinfo{journal}{Combustion and Flame} \bibinfo{volume}{198}
  (\bibinfo{year}{2018}) \bibinfo{pages}{186--204}.
  \DOIprefix\doi{10.1016/j.combustflame.2018.09.008}.
\bibitem[{Hansen and Sutherland(2018)}]{HANSEN2018257}
\bibinfo{author}{M.~A. Hansen}, \bibinfo{author}{J.~C. Sutherland},
\newblock \bibinfo{title}{On the consistency of state vectors and {Jacobian}
  matrices},
\newblock \bibinfo{journal}{Combust. Flame} \bibinfo{volume}{193}
  (\bibinfo{year}{2018}) \bibinfo{pages}{257--271}.
  \DOIprefix\doi{j.combustflame.2018.03.017}.
\bibitem[{Spafford et~al.(2010)Spafford, Meredith, Vetter, Chen, Grout, and
  Sankaran}]{Spafford:2010aa}
\bibinfo{author}{K.~Spafford}, \bibinfo{author}{J.~Meredith},
  \bibinfo{author}{J.~Vetter}, \bibinfo{author}{J.~Chen},
  \bibinfo{author}{R.~Grout}, \bibinfo{author}{R.~Sankaran},
\newblock \bibinfo{title}{Accelerating {S3D}: A {GPGPU} case study},
\newblock in: \bibinfo{booktitle}{Euro-Par 2009 Parallel Process. Workshops,
  LNCS 6043}, \bibinfo{publisher}{Springer-Verlag}, \bibinfo{address}{Berlin,
  Heidelberg}, \bibinfo{year}{2010}, pp. \bibinfo{pages}{122--131}.
  \DOIprefix\doi{10.1007/978-3-642-14122-5_16}.
\bibitem[{Shi et~al.(2011)Shi, Green, Wong, and Oluwole}]{Shi:2011aa}
\bibinfo{author}{Y.~Shi}, \bibinfo{author}{W.~H. Green}, \bibinfo{author}{H.-W.
  Wong}, \bibinfo{author}{O.~O. Oluwole},
\newblock \bibinfo{title}{Redesigning combustion modeling algorithms for the
  graphics processing unit ({GPU}): Chemical kinetic rate evaluation and
  ordinary differential equation integration},
\newblock \bibinfo{journal}{Combust. Flame} \bibinfo{volume}{158}
  (\bibinfo{year}{2011}) \bibinfo{pages}{836--847}.
  \DOIprefix\doi{10.1016/j.combustflame.2011.01.024}.
\bibitem[{Niemeyer et~al.(2011)Niemeyer, Sung, Fotache, and
  Lee}]{Niemeyer:2011aa}
\bibinfo{author}{K.~E. Niemeyer}, \bibinfo{author}{C.~J. Sung},
  \bibinfo{author}{C.~G. Fotache}, \bibinfo{author}{J.~C. Lee},
\newblock \bibinfo{title}{Turbulence-chemistry closure method using graphics
  processing units: a preliminary test},
\newblock in: \bibinfo{booktitle}{Fall 2011 Technical Meeting of the Eastern
  States Section of the Combust. Institute}, \bibinfo{year}{2011}.
  \DOIprefix\doi{10.6084/m9.figshare.3384964}.
\bibitem[{Stone and Davis(2013)}]{Stone:2013aa}
\bibinfo{author}{C.~P. Stone}, \bibinfo{author}{R.~L. Davis},
\newblock \bibinfo{title}{Techniques for solving stiff chemical kinetics on
  graphical processing units},
\newblock \bibinfo{journal}{J. Propul. Power} \bibinfo{volume}{29}
  (\bibinfo{year}{2013}) \bibinfo{pages}{764--773}.
  \DOIprefix\doi{10.2514/1.B34874}.
\bibitem[{Sewerin and Rigopoulos(2015)}]{Sewerin20151375}
\bibinfo{author}{F.~Sewerin}, \bibinfo{author}{S.~Rigopoulos},
\newblock \bibinfo{title}{A methodology for the integration of stiff chemical
  kinetics on {GPUs}},
\newblock \bibinfo{journal}{Combust. Flame} \bibinfo{volume}{162}
  (\bibinfo{year}{2015}) \bibinfo{pages}{1375--1394}.
  \DOIprefix\doi{10.1016/j.combustflame.2014.11.003}.
\bibitem[{Curtis et~al.(2017)Curtis, Niemeyer, and
  Sung}]{curtis2017investigation}
\bibinfo{author}{N.~J. Curtis}, \bibinfo{author}{K.~E. Niemeyer},
  \bibinfo{author}{C.~J. Sung},
\newblock \bibinfo{title}{An investigation of {GPU-based} stiff chemical
  kinetics integration methods},
\newblock \bibinfo{journal}{Combustion and Flame} \bibinfo{volume}{179}
  (\bibinfo{year}{2017}) \bibinfo{pages}{312--324}.
  \DOIprefix\doi{10.1016/j.combustflame.2017.02.005}.
\bibitem[{Kroshko and Spiteri(2013)}]{kroshko2013efficient}
\bibinfo{author}{A.~Kroshko}, \bibinfo{author}{R.~J. Spiteri},
\newblock \bibinfo{title}{Efficient {SIMD} solution of multiple systems of
  stiff {IVPs}},
\newblock \bibinfo{journal}{J. Comput. Sci} \bibinfo{volume}{4}
  (\bibinfo{year}{2013}) \bibinfo{pages}{377--385}.
  \DOIprefix\doi{10.1016/j.jocs.2012.08.017}.
\bibitem[{Linford and Sandu(2009)}]{Linford:2009}
\bibinfo{author}{J.~C. Linford}, \bibinfo{author}{A.~Sandu},
\newblock \bibinfo{title}{Chemical kinetics on multi-core {SIMD}
  architectures},
\newblock in: \bibinfo{booktitle}{Proc. 9th Int. Conf. Comput. Sci.},
  \bibinfo{year}{2009}.
\bibitem[{Linford et~al.(2011)Linford, Michalakes, Vachharajani, and
  Sandu}]{Linford:2011}
\bibinfo{author}{J.~C. Linford}, \bibinfo{author}{J.~Michalakes},
  \bibinfo{author}{M.~Vachharajani}, \bibinfo{author}{A.~Sandu},
\newblock \bibinfo{title}{Automatic generation of multicore chemical kernels},
\newblock \bibinfo{journal}{IEEE Trans. Parallel Distrib. Syst.}
  \bibinfo{volume}{22} (\bibinfo{year}{2011}) \bibinfo{pages}{119--131}.
  \DOIprefix\doi{10.1109/TPDS.2010.106}.
\bibitem[{Che et~al.(2018)Che, Yang, Xu, and Lu}]{CHE2018101}
\bibinfo{author}{Y.~Che}, \bibinfo{author}{M.~Yang}, \bibinfo{author}{C.~Xu},
  \bibinfo{author}{Y.~Lu},
\newblock \bibinfo{title}{Petascale scramjet combustion simulation on the
  {Tianhe-2} heterogeneous supercomputer},
\newblock \bibinfo{journal}{Parallel Comput.} \bibinfo{volume}{77}
  (\bibinfo{year}{2018}) \bibinfo{pages}{101--117}.
  \DOIprefix\doi{10.1016/j.parco.2018.06.004}.
\bibitem[{Dagum and Menon(1998)}]{dagum1998openmp}
\bibinfo{author}{L.~Dagum}, \bibinfo{author}{R.~Menon},
\newblock \bibinfo{title}{{OpenMP}: an industry standard {API} for
  shared-memory programming},
\newblock \bibinfo{journal}{Comput. Sci. \& Engineering, IEEE}
  \bibinfo{volume}{5} (\bibinfo{year}{1998}) \bibinfo{pages}{46--55}.
  \DOIprefix\doi{10.1109/99.660313}.
\bibitem[{Stone et~al.(2010)Stone, Gohara, and Shi}]{stone2010opencl}
\bibinfo{author}{J.~E. Stone}, \bibinfo{author}{D.~Gohara},
  \bibinfo{author}{G.~Shi},
\newblock \bibinfo{title}{{OpenCL}: A parallel programming standard for
  heterogeneous computing systems},
\newblock \bibinfo{journal}{IEEE Des. Test} \bibinfo{volume}{12}
  (\bibinfo{year}{2010}) \bibinfo{pages}{66--73}.
  \DOIprefix\doi{10.1109/MCSE.2010.69}.
\bibitem[{Hairer and Wanner(1996)}]{wanner1991solving}
\bibinfo{author}{E.~Hairer}, \bibinfo{author}{G.~Wanner},
  \bibinfo{title}{Solving Ordinary Differential Equations II: Stiff and
  Differential-Algebraic Problems}, \bibinfo{edition}{2} ed.,
  \bibinfo{publisher}{Springer-Verlag, Berlin}, \bibinfo{year}{1996}.
  \DOIprefix\doi{10.1007/978-3-642-05221-7}.
\bibitem[{Weller et~al.(1998)Weller, Tabor, Jasak, and Fureby}]{openfoam_paper}
\bibinfo{author}{H.~G. Weller}, \bibinfo{author}{G.~Tabor},
  \bibinfo{author}{H.~Jasak}, \bibinfo{author}{C.~Fureby},
\newblock \bibinfo{title}{A tensorial approach to computational continuum
  mechanics using object-oriented techniques},
\newblock \bibinfo{journal}{Comput. Phys.} \bibinfo{volume}{12}
  (\bibinfo{year}{1998}) \bibinfo{pages}{620--631}.
  \DOIprefix\doi{10.1063/1.168744}.
\bibitem[{Kee et~al.(1986)Kee, Dixon-Lewis, Warnatz, Coltrin, and
  Miller}]{kee1986fortran}
\bibinfo{author}{R.~J. Kee}, \bibinfo{author}{G.~Dixon-Lewis},
  \bibinfo{author}{J.~Warnatz}, \bibinfo{author}{M.~E. Coltrin},
  \bibinfo{author}{J.~A. Miller}, \bibinfo{title}{A {Fortran} computer code
  package for the evaluation of gas-phase multicomponent transport properties},
  \bibinfo{type}{Technical Report}, Sandia National Laboratories Report
  SAND86-8246, \bibinfo{year}{1986}.
\bibitem[{Baum et~al.(1995)Baum, Poinsot, and Th{\'e}venin}]{BAUM1995247}
\bibinfo{author}{M.~Baum}, \bibinfo{author}{T.~Poinsot},
  \bibinfo{author}{D.~Th{\'e}venin},
\newblock \bibinfo{title}{Accurate boundary conditions for multicomponent
  reactive flows},
\newblock \bibinfo{journal}{J. Comput. Phys.} \bibinfo{volume}{116}
  (\bibinfo{year}{1995}) \bibinfo{pages}{247 -- 261}.
  \DOIprefix\doi{10.1006/jcph.1995.1024}.
\bibitem[{Okong'o and Bellan(2002)}]{OKONGO2002330}
\bibinfo{author}{N.~Okong'o}, \bibinfo{author}{J.~Bellan},
\newblock \bibinfo{title}{Consistent boundary conditions for multicomponent
  real gas mixtures based on characteristic waves},
\newblock \bibinfo{journal}{J. Comput. Phys.} \bibinfo{volume}{176}
  (\bibinfo{year}{2002}) \bibinfo{pages}{330 -- 344}.
  \DOIprefix\doi{10.1006/jcph.2002.6990}.
\bibitem[{Cocks et~al.(2015)Cocks, Soteriou, and Sankaran}]{COCKS20153394}
\bibinfo{author}{P.~A. Cocks}, \bibinfo{author}{M.~C. Soteriou},
  \bibinfo{author}{V.~Sankaran},
\newblock \bibinfo{title}{Impact of numerics on the predictive capabilities of
  reacting flow {LES}},
\newblock \bibinfo{journal}{Combust. Flame} \bibinfo{volume}{162}
  (\bibinfo{year}{2015}) \bibinfo{pages}{3394--3411}.
  \DOIprefix\doi{10.1016/j.combustflame.2015.04.016}.
\bibitem[{Sjunnesson et~al.(1991{\natexlab{a}})Sjunnesson, Nelsson, and
  Max}]{sjunnesson1991lda}
\bibinfo{author}{A.~Sjunnesson}, \bibinfo{author}{C.~Nelsson},
  \bibinfo{author}{E.~Max},
\newblock \bibinfo{title}{{LDA} measurements of velocities and turbulence in a
  bluff body stabilized flame},
\newblock in: \bibinfo{booktitle}{Fourth International Conference on Laser
  Anemometry -- Advances and Application}, volume~\bibinfo{volume}{3},
  \bibinfo{address}{Cleveland, OH}, \bibinfo{year}{1991}{\natexlab{a}}, pp.
  \bibinfo{pages}{83--90}.
\bibitem[{Sjunnesson et~al.(1991{\natexlab{b}})Sjunnesson, Olovsson, and
  Sjoblom}]{sjunnesson1991validation}
\bibinfo{author}{A.~Sjunnesson}, \bibinfo{author}{S.~Olovsson},
  \bibinfo{author}{B.~Sjoblom},
\newblock \bibinfo{title}{Validation rig - a tool for flame studies},
\newblock in: \bibinfo{booktitle}{International Symposium on Air Breathing
  Engines, 10th, Nottingham, England}, \bibinfo{year}{1991}{\natexlab{b}}, pp.
  \bibinfo{pages}{385--393}.
\bibitem[{Sjunnesson et~al.(1992)Sjunnesson, Henrikson, and
  Lofstrom}]{sjunnesson1992cars}
\bibinfo{author}{A.~Sjunnesson}, \bibinfo{author}{P.~Henrikson},
  \bibinfo{author}{C.~Lofstrom},
\newblock \bibinfo{title}{{CARS} measurements and visualization of reacting
  flows in a bluff body stabilized flame},
\newblock in: \bibinfo{booktitle}{28th Joint Propulsion Conference and
  Exhibit}, \bibinfo{publisher}{AIAA}, \bibinfo{year}{1992}.
  \DOIprefix\doi{10.2514/6.1992-3650}, \bibinfo{note}{{AIAA} Paper No.
  92-3650}.
\bibitem[{Rochette et~al.(2018)Rochette, Collin-Bastiani, Gicquel, Vermorel,
  Veynante, and Poinsot}]{ROCHETTE2018417}
\bibinfo{author}{B.~Rochette}, \bibinfo{author}{F.~Collin-Bastiani},
  \bibinfo{author}{L.~Gicquel}, \bibinfo{author}{O.~Vermorel},
  \bibinfo{author}{D.~Veynante}, \bibinfo{author}{T.~Poinsot},
\newblock \bibinfo{title}{Influence of chemical schemes, numerical method and
  dynamic turbulent combustion modeling on {LES} of premixed turbulent flames},
\newblock \bibinfo{journal}{Combust. Flame} \bibinfo{volume}{191}
  (\bibinfo{year}{2018}) \bibinfo{pages}{417--430}.
  \DOIprefix\doi{10.1016/j.combustflame.2018.01.016}.
\bibitem[{Barlow and Frank(1998)}]{BARLOW19981087}
\bibinfo{author}{R.~Barlow}, \bibinfo{author}{J.~Frank},
\newblock \bibinfo{title}{Effects of turbulence on species mass fractions in
  methane/air jet flames},
\newblock \bibinfo{journal}{Twenty-Seventh International Symposium on
  Combustion} \bibinfo{volume}{27} (\bibinfo{year}{1998})
  \bibinfo{pages}{1087--1095}. \DOIprefix\doi{10.1016/S0082-0784(98)80510-9}.
\bibitem[{Barlow et~al.(2005)Barlow, Frank, Karpetis, and Chen}]{BARLOW2005433}
\bibinfo{author}{R.~Barlow}, \bibinfo{author}{J.~Frank},
  \bibinfo{author}{A.~Karpetis}, \bibinfo{author}{J.-Y. Chen},
\newblock \bibinfo{title}{Piloted methane/air jet flames: Transport effects and
  aspects of scalar structure},
\newblock \bibinfo{journal}{Combust. Flame} \bibinfo{volume}{143}
  (\bibinfo{year}{2005}) \bibinfo{pages}{433--449}.
  \DOIprefix\doi{https://doi.org/10.1016/j.combustflame.2005.08.017}.
\bibitem[{Schneider et~al.(2003)Schneider, Dreizler, Janicka, and
  Hassel}]{SCHNEIDER2003185}
\bibinfo{author}{C.~Schneider}, \bibinfo{author}{A.~Dreizler},
  \bibinfo{author}{J.~Janicka}, \bibinfo{author}{E.~Hassel},
\newblock \bibinfo{title}{Flow field measurements of stable and locally
  extinguishing hydrocarbon-fuelled jet flames},
\newblock \bibinfo{journal}{Combust. Flame} \bibinfo{volume}{135}
  (\bibinfo{year}{2003}) \bibinfo{pages}{185--190}.
  \DOIprefix\doi{doi.org/10.1016/S0010-2180(03)00150-0}.
\bibitem[{Bauer et~al.(2014)Bauer, Treichler, and Aiken}]{Bauer:2014}
\bibinfo{author}{M.~Bauer}, \bibinfo{author}{S.~Treichler},
  \bibinfo{author}{A.~Aiken},
\newblock \bibinfo{title}{Singe: Leveraging warp specialization for high
  performance on {GPUs}},
\newblock \bibinfo{journal}{SIGPLAN Not.} \bibinfo{volume}{49}
  (\bibinfo{year}{2014}) \bibinfo{pages}{119--130}.
  \DOIprefix\doi{10.1145/2692916.2555258}.
\bibitem[{{\relax{Intel}\textregistered\
  {Corporation}}(2018)}]{intelopencl:2018}
\bibinfo{author}{{\relax{Intel}\textregistered\ {Corporation}}},
  \bibinfo{title}{{OpenCL}\texttrademark\ drivers and runtimes for
  {Intel}\textregistered\ architecture},
  \bibinfo{howpublished}{\url{https://software.intel.com/en-us/articles/opencl-drivers\#latest_CPU_runtime}},
  \bibinfo{year}{2018}.
\bibitem[{Sandu et~al.(1997)Sandu, Verwer, Blom, Spee, Carmichael, and
  Potra}]{SANDU19973459}
\bibinfo{author}{A.~Sandu}, \bibinfo{author}{J.~Verwer},
  \bibinfo{author}{J.~Blom}, \bibinfo{author}{E.~Spee},
  \bibinfo{author}{G.~Carmichael}, \bibinfo{author}{F.~Potra},
\newblock \bibinfo{title}{Benchmarking stiff {ODE} solvers for atmospheric
  chemistry problems {II}: {Rosenbrock} solvers},
\newblock \bibinfo{journal}{Atmospheric Environ.} \bibinfo{volume}{31}
  (\bibinfo{year}{1997}) \bibinfo{pages}{3459--3472}.
  \DOIprefix\doi{10.1016/S1352-2310(97)83212-8}.
\bibitem[{Zhang and Sandu(2014)}]{FATODE}
\bibinfo{author}{H.~Zhang}, \bibinfo{author}{A.~Sandu},
\newblock \bibinfo{title}{{FATODE}: A library for forward, adjoint, and tangent
  linear integration of odes},
\newblock \bibinfo{journal}{SIAM J. Sci. Comput.} \bibinfo{volume}{36}
  (\bibinfo{year}{2014}) \bibinfo{pages}{C504--C523}.
  \DOIprefix\doi{10.1137/130912335}.
\bibitem[{Hairer et~al.(1993)Hairer, N{\o}rsett, and Wanner}]{Hairer1993}
\bibinfo{author}{E.~Hairer}, \bibinfo{author}{S.~P. N{\o}rsett},
  \bibinfo{author}{G.~Wanner}, \bibinfo{title}{Solving Ordinary Differential
  Equations I: Nonstiff Problems}, \bibinfo{publisher}{Springer Berlin
  Heidelberg}, \bibinfo{address}{Berlin, Heidelberg}, \bibinfo{year}{1993}.
  \DOIprefix\doi{10.1007/978-3-540-78862-1}.
\bibitem[{Kaps et~al.(1985)Kaps, Poon, and Bui}]{Kaps1985}
\bibinfo{author}{P.~Kaps}, \bibinfo{author}{S.~W.~H. Poon},
  \bibinfo{author}{T.~D. Bui},
\newblock \bibinfo{title}{{Rosenbrock} methods for stiff {ODEs}: A comparison
  of {Richardson} extrapolation and embedding technique},
\newblock \bibinfo{journal}{Comput.} \bibinfo{volume}{34}
  (\bibinfo{year}{1985}) \bibinfo{pages}{17--40}.
  \DOIprefix\doi{10.1007/BF02242171}.
\bibitem[{Shampine(1982)}]{Shampine:1982:IRM:355993.355994}
\bibinfo{author}{L.~F. Shampine},
\newblock \bibinfo{title}{Implementation of {Rosenbrock} methods},
\newblock \bibinfo{journal}{ACM Trans. Math. Softw.} \bibinfo{volume}{8}
  (\bibinfo{year}{1982}) \bibinfo{pages}{93--113}.
  \DOIprefix\doi{10.1145/355993.355994}.
\bibitem[{Lysenko et~al.(2014)Lysenko, Ertesv{\aa}g, and Rian}]{Lysenko2014b}
\bibinfo{author}{D.~A. Lysenko}, \bibinfo{author}{I.~S. Ertesv{\aa}g},
  \bibinfo{author}{K.~E. Rian},
\newblock \bibinfo{title}{Numerical simulation of non-premixed turbulent
  combustion using the eddy dissipation concept and comparing with the steady
  laminar flamelet model},
\newblock \bibinfo{journal}{Flow, Turbul. Combust.} \bibinfo{volume}{93}
  (\bibinfo{year}{2014}) \bibinfo{pages}{577--605}.
  \DOIprefix\doi{10.1007/s10494-014-9551-7}.
\bibitem[{Fureby(2009)}]{Fureby2957}
\bibinfo{author}{C.~Fureby},
\newblock \bibinfo{journal}{Philos. Trans. Royal Soc. A: Math., Phys. Eng.
  Sci.} \bibinfo{volume}{367} (\bibinfo{year}{2009})
  \bibinfo{pages}{2957--2969}. \DOIprefix\doi{10.1098/rsta.2008.0271}.
\bibitem[{{The OpenFOAM Foundation}(2018)}]{openfoam_manual}
\bibinfo{author}{{The OpenFOAM Foundation}}, \bibinfo{title}{{OpenFOAM} v6 user
  guide}, \bibinfo{year}{2018}. \URLprefix
  \url{https://cfd.direct/openfoam/user-guide-v6/}, \bibinfo{note}{[Online;
  accessed 12/02/18]}.
\bibitem[{{Message Passing Interface Forum}(2012)}]{MPI3}
\bibinfo{author}{{Message Passing Interface Forum}}, \bibinfo{title}{{MPI}: A
  message-passing interface standard, version 3.0},
  \bibinfo{howpublished}{\url{https://www.mpi-forum.org/docs/mpi-3.0/mpi30-report.pdf}},
  \bibinfo{year}{2012}.
\bibitem[{Weller et~al.(1998)Weller, Tabor, Gosman, and Fureby}]{WELLER1998899}
\bibinfo{author}{H.~Weller}, \bibinfo{author}{G.~Tabor},
  \bibinfo{author}{A.~Gosman}, \bibinfo{author}{C.~Fureby},
\newblock \bibinfo{title}{Application of a flame-wrinkling {LES} combustion
  model to a turbulent mixing layer},
\newblock volume~\bibinfo{volume}{27}, \bibinfo{year}{1998}, pp.
  \bibinfo{pages}{899--907}. \DOIprefix\doi{10.1016/S0082-0784(98)80487-6}.
\bibitem[{Magnussen(2005)}]{magnussen2005eddy}
\bibinfo{author}{B.~F. Magnussen},
\newblock \bibinfo{title}{The eddy dissipation concept--a bridge between
  science and technology},
\newblock in: \bibinfo{booktitle}{ECCOMAS thematic conference on computational
  combustion}, \bibinfo{year}{2005}, pp. \bibinfo{pages}{21--24}.
\bibitem[{B{\"o}senhofer et~al.(2018)B{\"o}senhofer, Wartha, Jordan, and
  Harasek}]{en11071902}
\bibinfo{author}{M.~B{\"o}senhofer}, \bibinfo{author}{E.-M. Wartha},
  \bibinfo{author}{C.~Jordan}, \bibinfo{author}{M.~Harasek},
\newblock \bibinfo{title}{The eddy dissipation concept---analysis of different
  fine structure treatments for classical combustion},
\newblock \bibinfo{journal}{Energies} \bibinfo{volume}{11}
  (\bibinfo{year}{2018}) \bibinfo{pages}{1902}.
  \DOIprefix\doi{10.3390/en11071902}.
\bibitem[{Evans et~al.(2015)Evans, Medwell, and Tian}]{edc1}
\bibinfo{author}{M.~J. Evans}, \bibinfo{author}{P.~R. Medwell},
  \bibinfo{author}{Z.~F. Tian},
\newblock \bibinfo{title}{Modeling lifted jet flames in a heated coflow using
  an optimized eddy dissipation concept model},
\newblock \bibinfo{journal}{Combust. Sci. Tech.} \bibinfo{volume}{187}
  (\bibinfo{year}{2015}) \bibinfo{pages}{1093--1109}.
  \DOIprefix\doi{10.1080/00102202.2014.1002836}.
\bibitem[{Li et~al.(2017)Li, Cuoci, Sadiki, and Parente}]{edc2}
\bibinfo{author}{Z.~Li}, \bibinfo{author}{A.~Cuoci},
  \bibinfo{author}{A.~Sadiki}, \bibinfo{author}{A.~Parente},
\newblock \bibinfo{title}{Comprehensive numerical study of the {Adelaide} jet
  in hot-coflow burner by means of {RANS} and detailed chemistry},
\newblock \bibinfo{journal}{Energy} \bibinfo{volume}{139}
  (\bibinfo{year}{2017}) \bibinfo{pages}{555--570}.
  \DOIprefix\doi{10.1016/j.energy.2017.07.132}.
\bibitem[{Magnussen(1989)}]{magnussen1989modeling}
\bibinfo{author}{B.~F. Magnussen},
\newblock \bibinfo{title}{Modeling of {NOx} and soot formation by the eddy
  dissipation concept},
\newblock in: \bibinfo{booktitle}{International Flame Research Foundation First
  Topic Oriented Technical Meeting}, \bibinfo{year}{1989}, pp.
  \bibinfo{pages}{17--19}.
\bibitem[{Banks et~al.(2016)Banks, Collier, Hindmarsh, Serban, and
  Woodward}]{sundials:2.7.0}
\bibinfo{author}{E.~Banks}, \bibinfo{author}{A.~M. Collier},
  \bibinfo{author}{A.~C. Hindmarsh}, \bibinfo{author}{R.~Serban},
  \bibinfo{author}{C.~S. Woodward}, \bibinfo{title}{\texttt{SUNDIALS} v2.7.0},
  \bibinfo{howpublished}{\url{http://computation.llnl.gov/projects/sundials-suite-nonlinear-differential-algebraic-equation-solvers/download/sundials-2.7.0.tar.gz}},
  \bibinfo{year}{2016}.
\bibitem[{Brown et~al.(1989)Brown, Byrne, and Hindmarsh}]{Brown:1989vl}
\bibinfo{author}{P.~N. Brown}, \bibinfo{author}{G.~D. Byrne},
  \bibinfo{author}{A.~C. Hindmarsh},
\newblock \bibinfo{title}{{VODE}: a variable-coefficient {ODE} solver},
\newblock \bibinfo{journal}{SIAM J. Sci. Stat. Comput.} \bibinfo{volume}{10}
  (\bibinfo{year}{1989}) \bibinfo{pages}{1038--1051}.
  \DOIprefix\doi{10.1137/0910062}.
\bibitem[{Hindmarsh et~al.(2005)Hindmarsh, Brown, Grant, Lee, Serban, Shumaker,
  and Woodward}]{Hindmarsh:2005}
\bibinfo{author}{A.~C. Hindmarsh}, \bibinfo{author}{P.~N. Brown},
  \bibinfo{author}{K.~E. Grant}, \bibinfo{author}{S.~L. Lee},
  \bibinfo{author}{R.~Serban}, \bibinfo{author}{D.~E. Shumaker},
  \bibinfo{author}{C.~S. Woodward},
\newblock \bibinfo{title}{{SUNDIALS}: Suite of nonlinear and
  differential/algebraic equation solvers},
\newblock \bibinfo{journal}{ACM Trans. Math. Softw.} \bibinfo{volume}{31}
  (\bibinfo{year}{2005}) \bibinfo{pages}{363--396}.
  \DOIprefix\doi{10.1145/1089014.1089020}.
\bibitem[{Goodwin et~al.(2018)Goodwin, Moffat, and Speth}]{cantera}
\bibinfo{author}{D.~G. Goodwin}, \bibinfo{author}{H.~K. Moffat},
  \bibinfo{author}{R.~L. Speth}, \bibinfo{title}{{Cantera}: An object-oriented
  software toolkit for chemical kinetics, thermodynamics, and transport
  processes}, \bibinfo{howpublished}{\url{http://www.cantera.org}},
  \bibinfo{year}{2018}. \DOIprefix\doi{10.5281/zenodo.1174508},
  \bibinfo{note}{version 2.4.0}.
\bibitem[{Curtis and Niemeyer(2016)}]{Curtis2016:ch4}
\bibinfo{author}{N.~J. Curtis}, \bibinfo{author}{K.~E. Niemeyer},
  \bibinfo{title}{\texttt{ch4\_pasr\_data.bin}},
  \bibinfo{howpublished}{Figshare}, \bibinfo{year}{2016}.
  \DOIprefix\doi{10.6084/m9.figshare.4007418.v2}.
\bibitem[{Smith et~al.(1999)Smith, Golden, Frenklach, Moriarty, Eiteneer,
  Goldenberg, Bowman, Hanson, Song, Gardiner, Lissianski, and
  Qin}]{smith_gri-mech_30}
\bibinfo{author}{G.~P. Smith}, \bibinfo{author}{D.~M. Golden},
  \bibinfo{author}{M.~Frenklach}, \bibinfo{author}{N.~W. Moriarty},
  \bibinfo{author}{B.~Eiteneer}, \bibinfo{author}{M.~Goldenberg},
  \bibinfo{author}{C.~T. Bowman}, \bibinfo{author}{R.~K. Hanson},
  \bibinfo{author}{S.~Song}, \bibinfo{author}{W.~C. Gardiner},
  \bibinfo{author}{V.~V. Lissianski}, \bibinfo{author}{Z.~Qin},
  \bibinfo{title}{{GRI}-{Mech} 3.0},
  \bibinfo{howpublished}{\url{http://www.me.berkeley.edu/gri_mech/}},
  \bibinfo{year}{1999}.
\bibitem[{{The OpenFOAM Foundation}(2017)}]{openfoam_foundation}
\bibinfo{author}{{The OpenFOAM Foundation}}, \bibinfo{title}{{OpenFOAM} v5.0},
  \bibinfo{year}{2017}. \URLprefix
  \url{https://openfoam.org/download/5-0-source/}, \bibinfo{note}{[Online;
  accessed 12/02/18]}.
\bibitem[{Gabriel et~al.(2004)Gabriel, Fagg, Bosilca, Angskun, Dongarra,
  Squyres, Sahay, Kambadur, Barrett, Lumsdaine, Castain, Daniel, Graham, and
  Woodall}]{open_mpi}
\bibinfo{author}{E.~Gabriel}, \bibinfo{author}{G.~E. Fagg},
  \bibinfo{author}{G.~Bosilca}, \bibinfo{author}{T.~Angskun},
  \bibinfo{author}{J.~J. Dongarra}, \bibinfo{author}{J.~M. Squyres},
  \bibinfo{author}{V.~Sahay}, \bibinfo{author}{P.~Kambadur},
  \bibinfo{author}{B.~Barrett}, \bibinfo{author}{A.~Lumsdaine},
  \bibinfo{author}{R.~H. Castain}, \bibinfo{author}{D.~J. Daniel},
  \bibinfo{author}{R.~L. Graham}, \bibinfo{author}{T.~S. Woodall},
\newblock \bibinfo{title}{Open {MPI}: Goals, concept, and design of a next
  generation {MPI} implementation},
\newblock in: \bibinfo{booktitle}{Recent Advances in Parallel Virtual Machine
  and Message Passing Interface}, \bibinfo{publisher}{Springer Berlin
  Heidelberg}, \bibinfo{year}{2004}, pp. \bibinfo{pages}{97--104}.
  \DOIprefix\doi{10.1007/978-3-540-30218-6_19}.
\bibitem[{Stallman and \relax{GCC Developer
  Community}(2009)}]{Stallman:2009:UGC:1593499}
\bibinfo{author}{R.~M. Stallman}, \bibinfo{author}{\relax{GCC Developer
  Community}}, \bibinfo{title}{Using The {GNU} Compiler Collection: A {GNU}
  Manual For {GCC} Version 4.3.3}, \bibinfo{publisher}{CreateSpace},
  \bibinfo{address}{Paramount, CA}, \bibinfo{year}{2009}.
\bibitem[{Skinner(2005)}]{skinner2005performance}
\bibinfo{author}{D.~Skinner}, \bibinfo{title}{Performance monitoring of
  parallel scientific applications}, \bibinfo{type}{Technical Report},
  \bibinfo{year}{2005}. \DOIprefix\doi{10.2172/881368}.
\bibitem[{Fureby and Moller(1995)}]{volvo1}
\bibinfo{author}{C.~Fureby}, \bibinfo{author}{S.-I. Moller},
\newblock \bibinfo{title}{Large eddy simulation of reacting flows applied to
  bluff body stabilized flames},
\newblock \bibinfo{journal}{{AIAA} J.} \bibinfo{volume}{33}
  (\bibinfo{year}{1995}) \bibinfo{pages}{2339--2347}.
  \DOIprefix\doi{10.2514/3.12989}.
\bibitem[{Zettervall et~al.(2017)Zettervall, Nordin-Bates, Nilsson, and
  Fureby}]{volvo2}
\bibinfo{author}{N.~Zettervall}, \bibinfo{author}{K.~Nordin-Bates},
  \bibinfo{author}{E.~Nilsson}, \bibinfo{author}{C.~Fureby},
\newblock \bibinfo{title}{Large eddy simulation of a premixed bluff body
  stabilized flame using global and skeletal reaction mechanisms},
\newblock \bibinfo{journal}{Combust. Flame} \bibinfo{volume}{179}
  (\bibinfo{year}{2017}) \bibinfo{pages}{1--22}.
  \DOIprefix\doi{10.1016/j.combustflame.2016.12.007}.
\bibitem[{M{\"o}ller et~al.(1996)M{\"o}ller, Lundgren, and Fureby}]{volvo3}
\bibinfo{author}{S.~M{\"o}ller}, \bibinfo{author}{E.~Lundgren},
  \bibinfo{author}{C.~Fureby},
\newblock \bibinfo{title}{Large eddy simulation of unsteady combustion},
\newblock \bibinfo{journal}{26th International Symposium on Combustion}
  \bibinfo{volume}{26} (\bibinfo{year}{1996}) \bibinfo{pages}{241--248}.
  \DOIprefix\doi{10.1016/S0082-0784(96)80222-0}.
\bibitem[{Lee and Cant(2017)}]{volvo4}
\bibinfo{author}{C.~Y. Lee}, \bibinfo{author}{S.~Cant},
\newblock \bibinfo{title}{Large-eddy simulation of a bluff-body stabilised
  turbulent premixed flame using the transported flame surface density
  approach},
\newblock \bibinfo{journal}{Combust. Theor. Model.} \bibinfo{volume}{21}
  (\bibinfo{year}{2017}) \bibinfo{pages}{722--748}.
  \DOIprefix\doi{10.1080/13647830.2017.1293849}.
\bibitem[{Comer(2018)}]{mvp}
\bibinfo{author}{A.~Comer}, \bibinfo{title}{3rd model validation for propulsion
  workshop overview and validation cases},
  \bibinfo{howpublished}{\url{https://community.apan.org/cfs-file/__key/widgetcontainerfiles/3fc3f82483d14ec485ef92e206116d49-g-_2D00_tM6tEO4PkenM5KsnY8ctg-page-0cases/MVP3_5F00_Validation_5F00_Case_5F00_20180608.pdf}},
  \bibinfo{year}{2018}. \bibinfo{note}{Accessed: 01-08-19}.
\bibitem[{Comer(2016)}]{mvp_exp}
\bibinfo{author}{A.~Comer}, \bibinfo{title}{Model validation for propulsion
  workshop - experimental data archives}, \bibinfo{year}{2016}. \URLprefix
  \url{https://community.apan.org/wg/afrlcg/mvpws/p/experimental-data},
  \bibinfo{note}{accessed: 01-08-19}.
\bibitem[{Smagorinsky(1963)}]{smag}
\bibinfo{author}{J.~Smagorinsky},
\newblock \bibinfo{title}{General circulation experiments with the primitive
  equations},
\newblock \bibinfo{journal}{Mon. Weather Review} \bibinfo{volume}{91}
  (\bibinfo{year}{1963}) \bibinfo{pages}{99--164}.
  \DOIprefix\doi{10.1175/1520-0493(1963)091<0099:GCEWTP>2.3.CO;2}.
\bibitem[{Issa(1986)}]{Issa1986}
\bibinfo{author}{R.~I. Issa},
\newblock \bibinfo{title}{Solution of the implicitly discretised fluid flow
  equations by operator-splitting},
\newblock \bibinfo{journal}{Journal of Computational Physics}
  \bibinfo{volume}{62} (\bibinfo{year}{1986}) \bibinfo{pages}{40--65}.
  \DOIprefix\doi{10.1016/0021-9991(86)90099-9}.
\bibitem[{Weller et~al.(2018)Weller, Greenshields, and Bainbridge}]{OFV6}
\bibinfo{author}{H.~Weller}, \bibinfo{author}{C.~Greenshields},
  \bibinfo{author}{W.~Bainbridge}, \bibinfo{title}{\texttt{OpenFOAM} v6},
  \bibinfo{howpublished}{\href{https://openfoam.org/release/6/}{https://openfoam.org/release/6/}},
  \bibinfo{year}{2018}. \bibinfo{note}{Note:
  \href{https://github.com/OpenFOAM/OpenFOAM-6/commit/00e347c45aa37317c0244803583a589078993fd2}{Commit
  00e347 on the \texttt{OpenFOAM-6} GitHub repository}}.
\bibitem[{{Mechanical and Aerospace Engineering (Combustion Research),
  University of California at San Diego}(2018)}]{ucsd}
\bibinfo{author}{{Mechanical and Aerospace Engineering (Combustion Research),
  University of California at San Diego}}, \bibinfo{title}{Chemical-kinetic
  mechanisms for combustion applications},
  \bibinfo{howpublished}{\url{http://combustion.ucsd.edu}},
  \bibinfo{year}{2018}. \bibinfo{note}{{San Diego Mechanism web page}}.
\bibitem[{Franzelli et~al.(2012)Franzelli, Riber, Gicquel, and
  Poinsot}]{FRANZELLI2012621}
\bibinfo{author}{B.~Franzelli}, \bibinfo{author}{E.~Riber},
  \bibinfo{author}{L.~Y. Gicquel}, \bibinfo{author}{T.~Poinsot},
\newblock \bibinfo{title}{Large eddy simulation of combustion instabilities in
  a lean partially premixed swirled flame},
\newblock \bibinfo{journal}{Combust. Flame} \bibinfo{volume}{159}
  (\bibinfo{year}{2012}) \bibinfo{pages}{621--637}.
  \DOIprefix\doi{10.1016/j.combustflame.2011.08.004}.
\bibitem[{Jones et~al.(01  )Jones, Oliphant, Peterson et~al.}]{scipy}
\bibinfo{author}{E.~Jones}, \bibinfo{author}{T.~Oliphant},
  \bibinfo{author}{P.~Peterson}, et~al., \bibinfo{title}{{SciPy}: Open source
  scientific tools for {Python}}, \bibinfo{year}{2001--}. \URLprefix
  \url{http://www.scipy.org/}, \bibinfo{note}{[Online; accessed 12/02/18]}.
\bibitem[{Virtanen et~al.(2020)Virtanen, Gommers, Oliphant, Haberland, Reddy,
  Cournapeau, Burovski, Peterson, Weckesser, Bright, van~der Walt, Brett,
  Wilson, Millman, Mayorov, Nelson, Jones, Kern, Larson, Carey, Polat, Feng,
  Moore, {VanderPlas}, Laxalde, Perktold, Cimrman, Henriksen, Quintero, Harris,
  Archibald, Ribeiro, Pedregosa, van Mulbregt, and {{SciPy 1.0
  Contributors}}}]{Virtanen2020}
\bibinfo{author}{P.~Virtanen}, \bibinfo{author}{R.~Gommers},
  \bibinfo{author}{T.~E. Oliphant}, \bibinfo{author}{M.~Haberland},
  \bibinfo{author}{T.~Reddy}, \bibinfo{author}{D.~Cournapeau},
  \bibinfo{author}{E.~Burovski}, \bibinfo{author}{P.~Peterson},
  \bibinfo{author}{W.~Weckesser}, \bibinfo{author}{J.~Bright},
  \bibinfo{author}{S.~J. van~der Walt}, \bibinfo{author}{M.~Brett},
  \bibinfo{author}{J.~Wilson}, \bibinfo{author}{K.~J. Millman},
  \bibinfo{author}{N.~Mayorov}, \bibinfo{author}{A.~R.~J. Nelson},
  \bibinfo{author}{E.~Jones}, \bibinfo{author}{R.~Kern},
  \bibinfo{author}{E.~Larson}, \bibinfo{author}{C.~J. Carey},
  \bibinfo{author}{{\.{I}}.~Polat}, \bibinfo{author}{Y.~Feng},
  \bibinfo{author}{E.~W. Moore}, \bibinfo{author}{J.~{VanderPlas}},
  \bibinfo{author}{D.~Laxalde}, \bibinfo{author}{J.~Perktold},
  \bibinfo{author}{R.~Cimrman}, \bibinfo{author}{I.~Henriksen},
  \bibinfo{author}{E.~A. Quintero}, \bibinfo{author}{C.~R. Harris},
  \bibinfo{author}{A.~M. Archibald}, \bibinfo{author}{A.~H. Ribeiro},
  \bibinfo{author}{F.~Pedregosa}, \bibinfo{author}{P.~van Mulbregt},
  \bibinfo{author}{{{SciPy 1.0 Contributors}}},
\newblock \bibinfo{title}{{SciPy} 1.0: fundamental algorithms for scientific
  computing in {Python}},
\newblock \bibinfo{journal}{Nature Methods} \bibinfo{volume}{17}
  (\bibinfo{year}{2020}) \bibinfo{pages}{261--272}.
  \DOIprefix\doi{10.1038/s41592-019-0686-2}.
\bibitem[{Kodavasal et~al.(2016)Kodavasal, Harms, Srivastava, Som, Quan,
  Richards, and Garc\'{i}a}]{kodavasal2016development}
\bibinfo{author}{J.~Kodavasal}, \bibinfo{author}{K.~Harms},
  \bibinfo{author}{P.~Srivastava}, \bibinfo{author}{S.~Som},
  \bibinfo{author}{S.~Quan}, \bibinfo{author}{K.~Richards},
  \bibinfo{author}{M.~Garc\'{i}a},
\newblock \bibinfo{title}{Development of a stiffness-based chemistry load
  balancing scheme, and optimization of input/output and communication, to
  enable massively parallel high-fidelity internal combustion engine
  simulations},
\newblock \bibinfo{journal}{J. Energy Resour. Technol.} \bibinfo{volume}{138}
  (\bibinfo{year}{2016}) \bibinfo{pages}{052203}.
  \DOIprefix\doi{10.1115/1.4032623}.
\bibitem[{{OpenMP Architecture Review Board}(2021)}]{openmp5}
\bibinfo{author}{{OpenMP Architecture Review Board}}, \bibinfo{title}{{OpenMP}
  {Application} {Program} {Interface} version 5.2}, \bibinfo{year}{2021}.
  \URLprefix
  \url{https://www.openmp.org/wp-content/uploads/OpenMP-API-Specification-5-2.pdf}.
\bibitem[{Curtis and Niemeyer(2019)}]{accelerInt:beta2}
\bibinfo{author}{N.~J. Curtis}, \bibinfo{author}{K.~E. Niemeyer},
  \bibinfo{title}{\texttt{accelerInt} v2.0-beta}, \bibinfo{year}{2019}.
  \URLprefix \url{https://github.com/SLACKHA/accelerInt/tree/rewrite}.
  \DOIprefix\doi{10.5281/zenodo.5963906}.
\bibitem[{Curtis(2019{\natexlab{a}})}]{bluffbody_les_case}
\bibinfo{author}{N.~J. Curtis}, \bibinfo{title}{\texttt{bluffbody\_LES}---an
  implementation of a {Volvo Flygmotor AB} bluff-body reacting flow simulation
  for {OpenFOAM} with a vectorized {ODE} solver},
  \bibinfo{howpublished}{\url{https://github.com/arghdos/bluffbody_LES}},
  \bibinfo{year}{2019}{\natexlab{a}}. \DOIprefix\doi{10.5281/zenodo.2574941}.
\bibitem[{Curtis(2019{\natexlab{b}})}]{reactingfoamipm}
\bibinfo{author}{N.~J. Curtis}, \bibinfo{title}{\texttt{reactingFoamIPM}-beta},
  \bibinfo{howpublished}{Zenodo [software], version \texttt{7b46091}},
  \bibinfo{year}{2019}{\natexlab{b}}. \URLprefix
  \url{https://github.com/arghdos/reactingFoamIPM}.
  \DOIprefix\doi{10.5281/zenodo.5963888}.
\bibitem[{Flowers et~al.(2006)Flowers, Aceves, and
  Babajimopoulos}]{Flowers2006}
\bibinfo{author}{D.~L. Flowers}, \bibinfo{author}{S.~M. Aceves},
  \bibinfo{author}{A.~Babajimopoulos},
\newblock \bibinfo{title}{Effect of charge non-uniformity on heat release and
  emissions in {PCCI} engine combustion},
\newblock in: \bibinfo{booktitle}{SAE Technical Paper Series},
  \bibinfo{number}{2006-01-1363}, \bibinfo{publisher}{SAE International},
  \bibinfo{year}{2006}. \DOIprefix\doi{10.4271/2006-01-1363}.
\bibitem[{Shi et~al.(2009)Shi, Kokjohn, Ge, and Reitz}]{Shi2009}
\bibinfo{author}{Y.~Shi}, \bibinfo{author}{S.~L. Kokjohn},
  \bibinfo{author}{H.-W. Ge}, \bibinfo{author}{R.~D. Reitz},
\newblock \bibinfo{title}{Efficient multidimensional simulation of {HCCI} and
  {DI} engine combustion with detailed chemistry},
\newblock in: \bibinfo{booktitle}{SAE Technical Paper Series},
  \bibinfo{number}{2009-01-0701}, \bibinfo{publisher}{SAE International},
  \bibinfo{year}{2009}. \DOIprefix\doi{10.4271/2009-01-0701}.
\bibitem[{Curtis and Niemeyer(2019)}]{pyJac_v2}
\bibinfo{author}{N.~J. Curtis}, \bibinfo{author}{K.~E. Niemeyer},
  \bibinfo{title}{\texttt{pyJac} v2.0}, \bibinfo{year}{2019}. \URLprefix
  \url{https://github.com/SLACKHA/pyJac-v2}.
  \DOIprefix\doi{10.5281/zenodo.5964176}.
\bibitem[{Curtis et~al.(2022)Curtis, Niemeyer, and Sung}]{paper_data}
\bibinfo{author}{N.~J. Curtis}, \bibinfo{author}{K.~E. Niemeyer},
  \bibinfo{author}{C.~J. Sung}, \bibinfo{title}{Data, plotting scripts, and
  figures for ``{Accelerating} reactive-flow simulations using vectorized
  chemistry integration''}, \bibinfo{howpublished}{Zenodo [dataset]},
  \bibinfo{year}{2022}. \DOIprefix\doi{10.5281/zenodo.5963861}.

\end{thebibliography}
\bibliographystyle{elsarticle-num-names}

\end{document}